\documentclass[a4paper,11pt]{article}
\usepackage{jinstpub}
\pdfoutput=1 

\usepackage{subfig}
\usepackage{diagbox}
\usepackage[squaren]{SIunits}

\title{Fast and Efficient Detection of 511~keV Photons using Cherenkov Light in PbF$_2$ Crystal, coupled to a MCP-PMT and SAMPIC Digitization Module.}

\author[a]{C.~Canot}
\author[a,c]{M.~Alokhina}
\author[a]{P.~Abbon}
\author[a]{J.P.~Bard}
\author[b]{D.~Breton} 
\author[a]{E.~Delagnes}
\author[b]{J.~Maalmi}
\author[a]{G.~Tauzin}
\author[a]{D.~Yvon}
\author[a,1]{and V.~Sharyy}\note{Corresponding author}

\affiliation[a]{IRFU, CEA,  Universit\'e Paris-Saclay,  Gif-sur-Yvette, France}
\affiliation[b]{LAL, IN2P3, CNRS, Universit\'e Paris-Saclay,Orsay, France}
\affiliation[c]{Taras Shevchenko National University of Kyiv, Kyiv, Ukraine}

\emailAdd{viatcheslav.sharyy@cea.fr}

\abstract{
We study the possibility to use the Cherenkov light for the efficient detection of 511~keV photons 
with the goal to use it in TOF-PET.
We designed and tested two detection modules consisting of PbF$_2$ crystals
attached to Planacon MCP-PMT XP85012. Amplified PMT signals are digitized by the 
SAMPIC module with high readout rate, up to ~$10^5$ events/s, and 
a negligible contribution to the time resolution, below 20~ps (FWHM).
We developed a fast 2D scanning system to calibrate the PMT time response and 
studied in details the timing characteristics of the Planacon PMT.

Using a radioactive $^{22}$Na source we  measured a detection efficiency of 24\% for 511 keV photons
in a 10 mm thick crystal and a coincidence resolving time of 280~ps.
We analyzed the main factors limiting the time resolution of the large-surface detection module and 
proposed solutions to improve it, which will be tested in our future project.
}

\keywords{Cherenkov Detector, Gamma Detector, Nuclear Imaging, PET, Time-Of-Flight, SAMPIC, Planacon, MCP-PMT}

\newcommand{\PbF} {PbF\ensuremath{_2}}

\newcommand{\Na}{\ensuremath{^{22}}Na}

\newcommand{\mm}{\milli\meter}
\newcommand{\um}{\micro\meter}
\newcommand{\db}{\deci\bel}
\newcommand{\ns}{\nano\second}

\begin{document}
\maketitle

\tableofcontents

\flushbottom

\section{Introduction}
Positron emission tomography (PET) is a nuclear imaging technique
widely used in oncology, cardiology,  neurobiology, preclinical researches~\cite{Kuntner2014, Walrand2016Nov}. 
PET technique has the ability to image and to quantify biochemical parameters by using
specially designed radiopharmaceuticals.
All PET tracers incorporate positron emitting, short-living isotopes like $^{18}F$, $^{15}O$, $^{11}C$ and others~\cite{Pike2009, Vallabhajosula_2011}. 
Annihilation of the emitted positron in the nearby tissue produces mainly two back-to-back 511~keV photons. Detection 
of both photons in coincidence allows  to reconstruct the line-of-response (LOR) and 
then image the tracer distribution in the object, by recording many coincidence events.

The quality of the image is determined to the large extent by the injected tracer activity and by the signal-to-background ratio.
Only several mm in the reconstructed LOR correspond to the annihilation vertex and all other LOR length contributes 
to the background counts in the reconstruction image.
The time-of-flight (TOF) technique allows to reduce efficiently the ``background'' length of LOR 
and increase the signal-to-background ratio. It consists in measuring the 
difference in the detection time of two 511~keV photons and using it as a spatial constraint
in the image reconstruction procedure. 
The TOF reconstruction produces images of better quality with a gain in the signal-to-noise  ratio, $G$,  
proportional to $\delta t$, the coincidence resolving time (CRT\footnote{Coincidence Resolving Time: 
width at half maximum of the time difference distribution.}).
More precisely, $G \simeq D/\Delta x$, where $D$ is a dimension 
of the object and $\Delta x$ is a localization uncertainty along LOR, 
defined as $\Delta x = c\delta t/2 $ with $c$ stands for the speed of light in the vacuum,
see e.g.~\cite{CAMPAGNOLO1979a, Tomitani1981, Yamamoto1982, Budinger1983, Conti_2011, Geramifar_2011, Westerwoudt_2014}. 

The TOF technique has been experimented in PET starting at yearly 80s 
\cite{Allemand1980Feb,Vacher1982,TerPogossian1982,Yamamoto1982,Bendriem1986,Trebossen1990Oct}, but only 
during the last decade, the commercial TOF PET scanners reach CRT of 
200 -- 500~ps~\cite{Surti2007a,Jakoby_2011,Bett2011,Zaidi2011Apr,Miller2015Jan, Huo2018Dec, vanSluis2019Jan}.
Laboratory studies using two detection modules obtain
much lower values of CRT, down to $\sim$100~ps for crystals with large thickness, 
20~mm typically \cite{Dam2013,Borghi_2016a,Berg2018,Cates2018Jun,Gundacker2019Feb}, 
and even 60~ps for crystals of 3~mm thickness~\cite{Gundacker2019Feb}.
All commercial devices and most of the experimental studies are done using the conventional technique, 
i.e. bright scintillator coupled to a photo-multiplier tube (PMT) or, recently, to the silicon photo-multiplier. 
At the same time, this approach is intrinsically limited by the scintillation signal rise and decay time.

Recently, an alternative detection method using the Cherenkov light received more  attention.
A relativistic electron created by the gamma quantum conversion emits Cherenkov photons at the time scale of several picoseconds, but
the number of the emitted photons is very limited, typically  10 -- 20. 
The possibility to  use the Cherenkov light in addition to the scintillation  and to improve the time resolution 
is discussed in~\cite{Lecoq_2010,Brunner_2013,Brunner_2014,Lecoq2017} and 
the improvement is measured experimentally using BGO crystals~\cite{Kwon_2016,Brunner2017} with
CRT values of 200~ps for crystal thickness 3~mm or 330~ps for 20~mm thickness.

Another type of study uses the pure Cherenkov radiator to characterize the possibility to detect 511 keV photons 
without using scintillation light. Promising results were obtained with lead glass~\cite{MIYATA_2006,Ota2019Mar}, liquid 
TMBi~\cite{Ramos2015} and crystalline \PbF~\cite{Korpar_2011,Korpar_2013,Dolenec2016Oct}.
In particular, small \PbF\ crystals coupled to a micro-channel-plate photo-multiplier tube (MCP-PMT) 
allow to reach a CRT of the order of 85~ps, but with a low detection efficiency of 8~\%~\cite{Korpar_2013}. 
The use of the lead glass as a PMT window allows to reach even better CRT of 30~ps, but 
the detection efficiency ``will not satisfy the requirement of clinical PET detector''~\cite{Ota2019Mar}.

In the presented study we investigate the possibility to create a large size PET Cherenkov
 detection module using a \PbF\ crystal coupled to a commercial 
MCP-PMT with a large detection efficiency, compatible with the use in PET scanner. 
We realized and tested two detection modules using fast digitizing electronics, able to provide a 
high data acquisition rate needed in clinical PET scans and tested their performance using a positrons emitting 
\Na source.

In the section~\ref{sec:det} we describe the detection module assembling and readout electronics, 
in the section~\ref{sec:eff} we discuss the detection efficiency measurements and in 
the section~\ref{sec:time} we present the time resolution study, including the detailed investigation 
of the PMT and readout electronics contribution.
The measured CRT and the possible improvements are discussed in the section~\ref{sec:CRT}.

\section{Detection Module and Readout Electronics}
\label{sec:det}

\begin{figure}
\begin{minipage}[t]{.49\textwidth} 
\includegraphics[width=\textwidth]{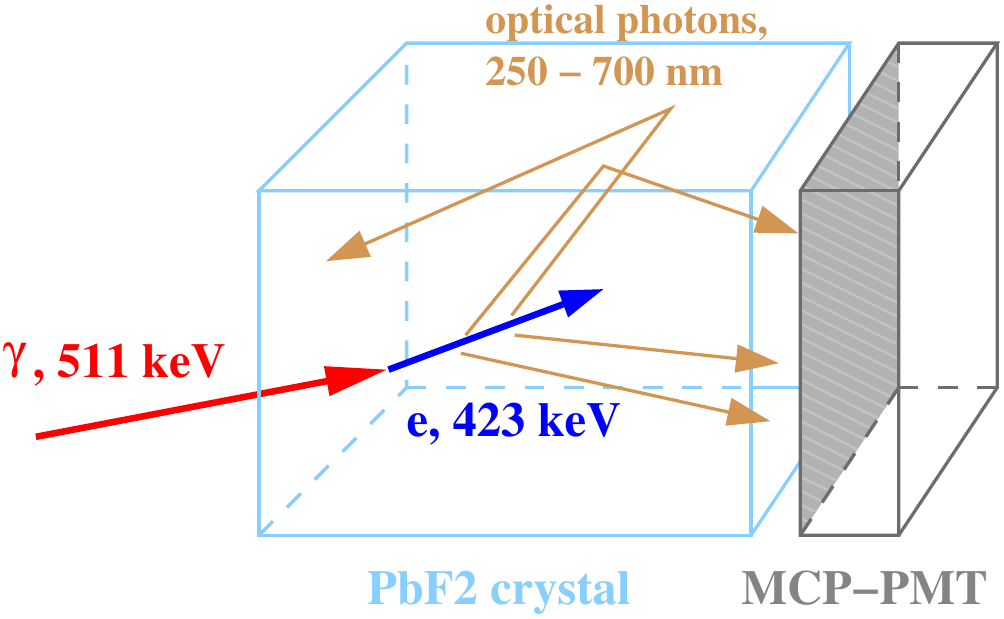}
\caption{Schematic view of the 511~keV photon detection module  with \PbF~crystal.
\label{fig:peche}}
\end{minipage}
\hfill
\begin{minipage}[t]{.49\textwidth} 
  \includegraphics[width=\textwidth]{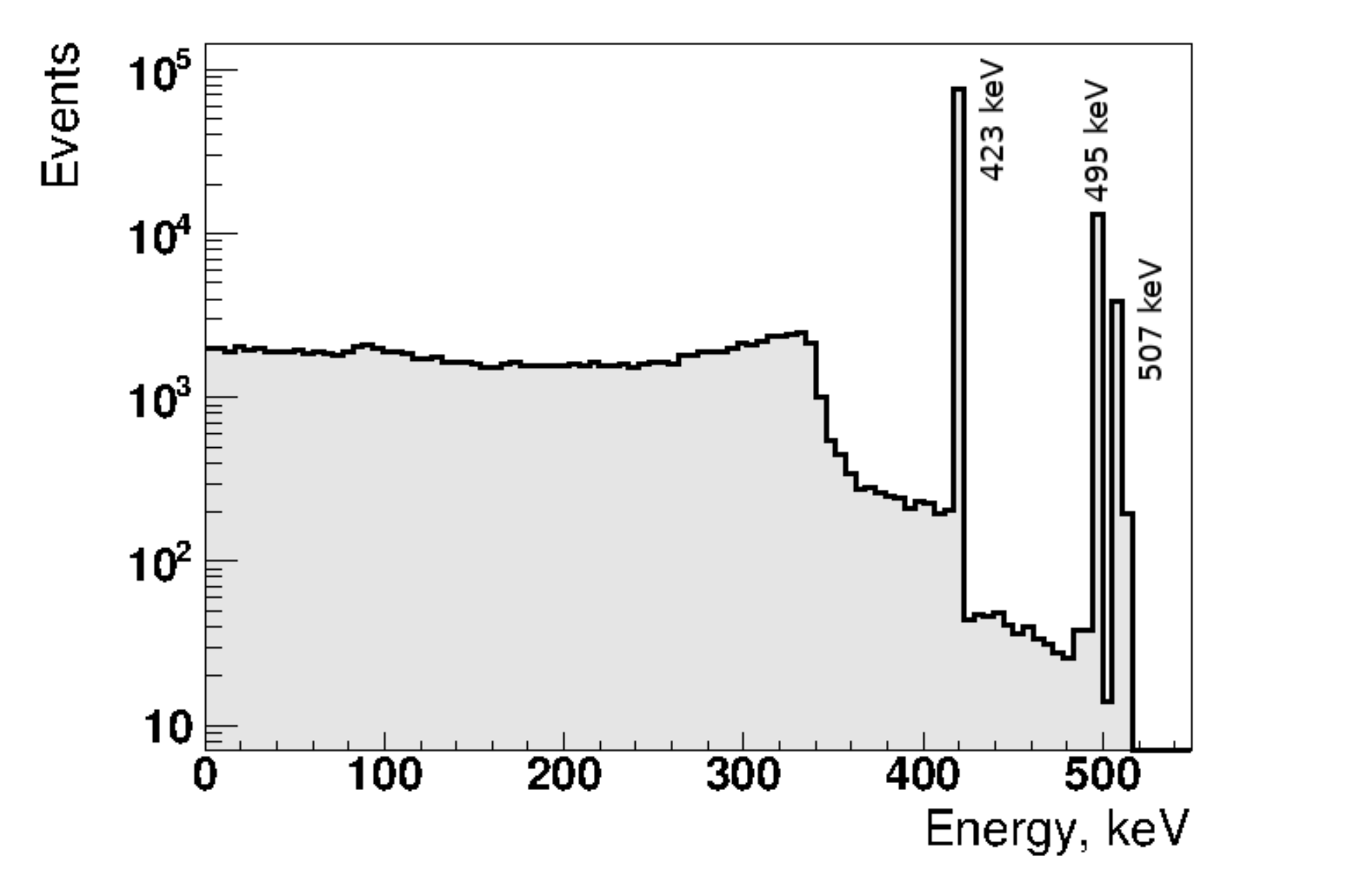}
  \caption{Simulated spectrum of the electron energy produced in the \unit{10}{\mm} \PbF~crystal by the 
     511~keV photon conversion. Three peaks correspond to the K, L and M-shell photoionization of lead.
    \label{fig:electron_energy}}
\end{minipage}
\end{figure}

Figure \ref{fig:peche} shows the schematic view of the 511~keV photon detection module, consisting of 
a monolithic  \PbF\ crystal with the size 53x53x10~mm$^3$ coupled to a MCP-PMT. 
To ensure the efficient light collection we use the optical gel OCF452 from Newgate \cite{OpticalGel} 
as an optical interface between crystal and PMT window.
The detection module is inserted in a black plastic packaging with the 10~mm wall thickness. 
PMT signals  are amplified by commercial 
amplifiers ZKL  and  digitized by the module SAMPIC, described later in the section~\ref{sec:readout}.  

\subsection{Cherenkov Radiator}
\label{sec:cherenkov_rad}
The  crystalline lead fluoride, \PbF, is a non-scintillating crystal transparent to the 
photon wavelength $\lambda>\unit{250}{\nano\meter}$~\cite{Williams1957,Anderson:1989uj}. 
Due to its large density, \unit{7.66}{\gram\per\centi\meter\cubed},  and  high atomic number of the lead 
it has a very short attenuation length of \unit{9}{\mm} for 511~keV photons.
When such a photon is converted in the crystal via the photoelectric effect (probability 46\% \cite{NIST}), 
it produces mainly a \unit{423}{keV} electron.
For the other 54\%, the photon is converted mainly via the Compton scattering process and produces an 
electron with energy less than \unit{340}{keV}, 
Fig.~\ref{fig:electron_energy}.
According to the Geant4~\cite{Agostinelli2003,Allison2006} simulation, the detection module with \unit{10}{\mm} 
crystal has a photon conversion probability of 75\%, where 30\% corresponds to events 
with a single photoionization conversion, 20\% to events with a single 
Compton scattering vertex and all other events have at least 
two vertices, e.g. one Compton and one photoionization vertices
 or two Compton vertices, etc. 
Conversion probability of 75\% is slightly higher than  expectation, 67\%, for the the 10~mm-thick crystal with the interaction length 9~mm.
This increase is due to the photons generated by the Compton scattering in materials surrounding the detector .

\begin{figure}
\begin{minipage}[t]{.49\textwidth}
  \includegraphics[width=\textwidth]{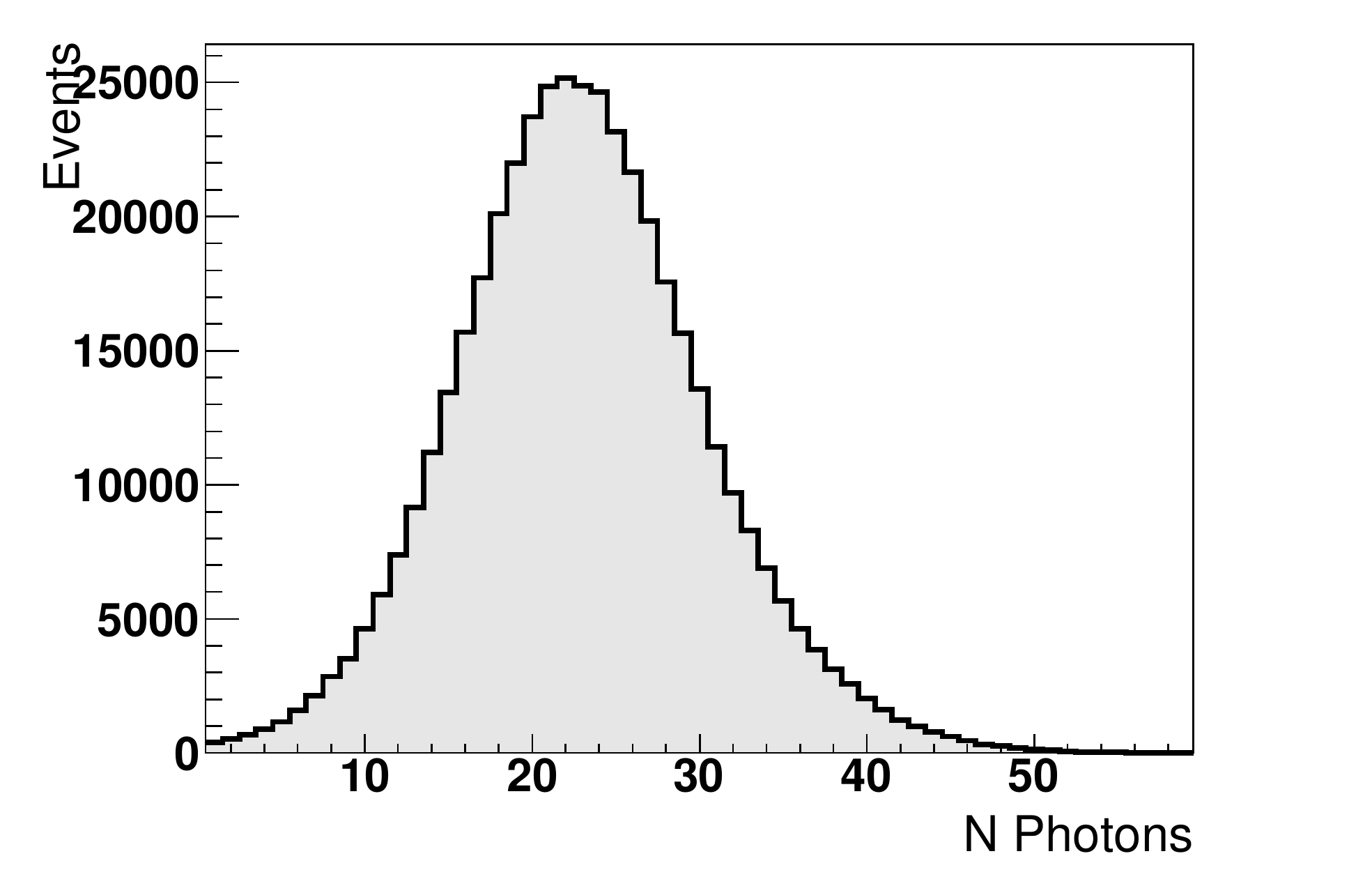}
  \caption{Simulated number of optical photons produced by electron, in the case of photoionization conversion of 511~keV photon.
    \label{fig:nphotons}}
\end{minipage}
\hfill
\begin{minipage}[t]{.49\textwidth} 
  \includegraphics[width=\textwidth]{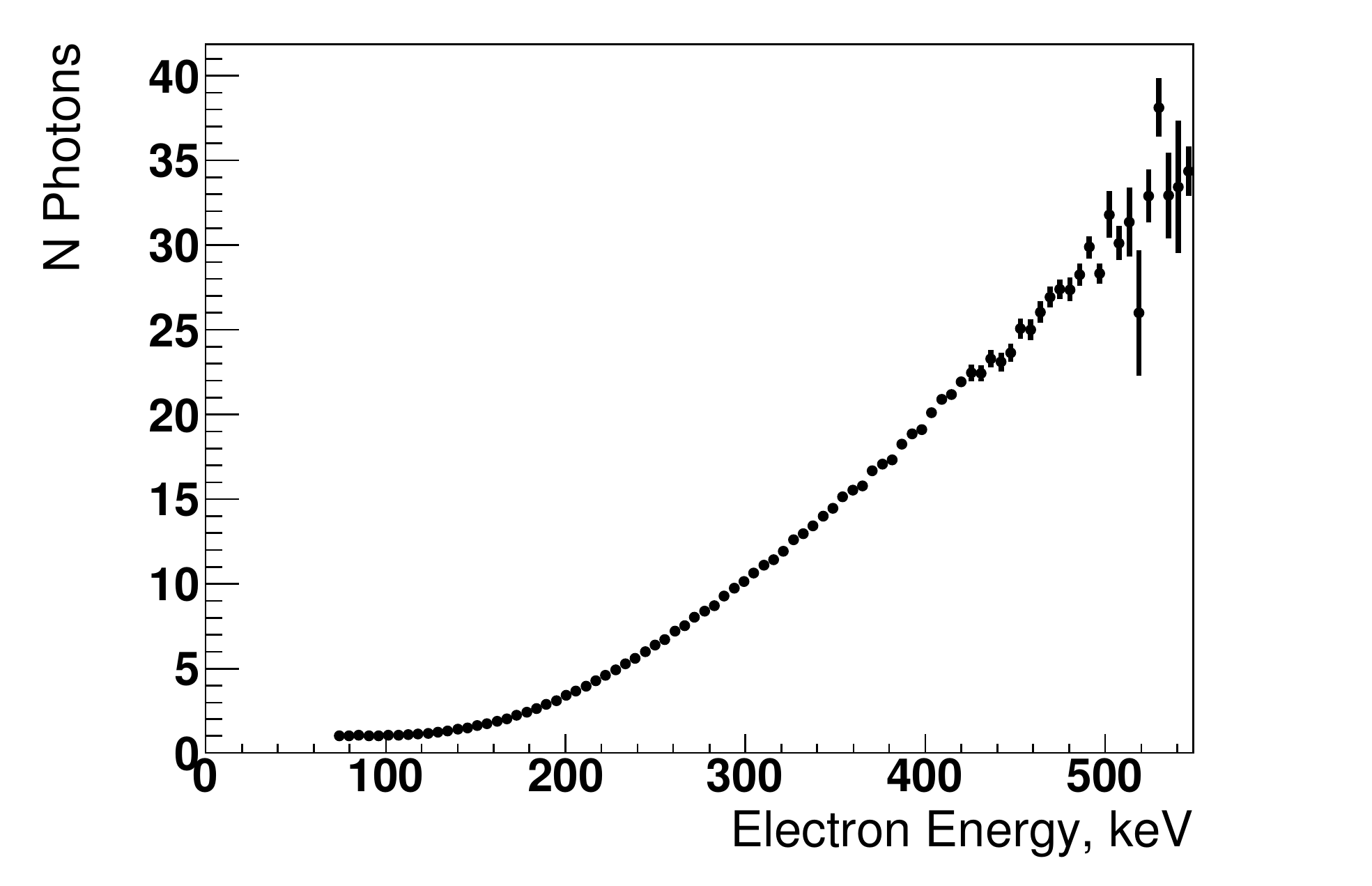}
  \caption{Mean number of Cherenkov photons produced by the electron in \PbF~crystal 
    versus the electron energy, according to the Geant4 simulation.
    \label{fig:nphotons_vs_energy}}
\end{minipage}
\end{figure}

The \PbF\ crystal is an excellent Cherenkov radiator due to the high refraction index of  1.82 at 400~nm wavelength.
Electrons emitted through the photoionization process  are sufficiently fast to produce
about 20 optical photons in average, Fig.~\ref{fig:nphotons}. Moreover, electrons from the Compton 
conversion  are also producing a smaller 
number of photons, as shown in Fig.~\ref{fig:nphotons_vs_energy}. They are contributing to the increase 
of the overall detection efficiency, 
but the detection efficiency of events through the Compton scattering  is smaller compared 
to events detected through the photoionization conversion.

\subsection{Photodetector}

\begin{figure}
  \centerline{\includegraphics[width=\textwidth]{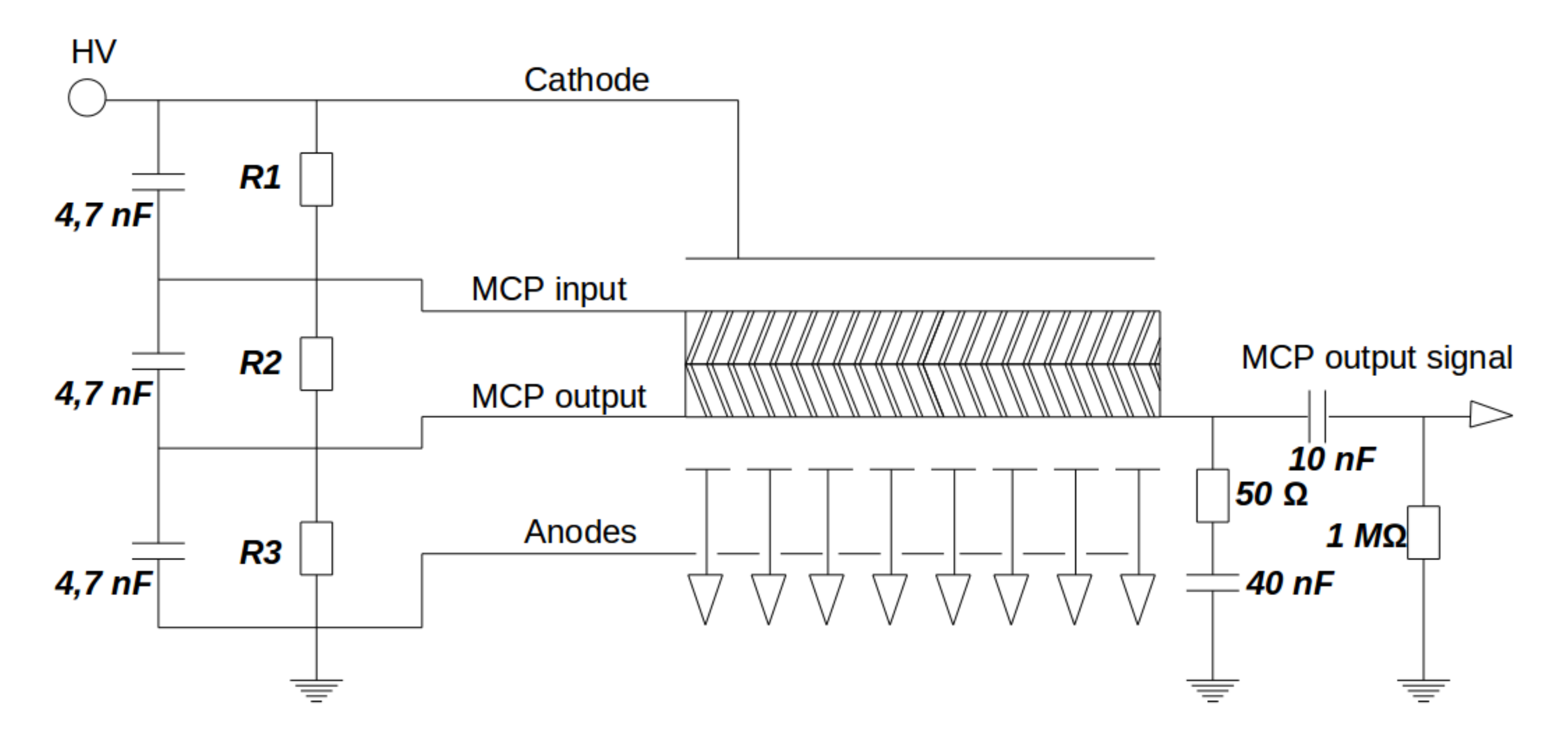}}
  \caption{Scheme of the PMT's high voltage divider and signal readout~\cite{XP85012_DataSheet}.  
    \label{fig:divider}}
\end{figure}

To detect optical  photons, we used a MCP-PMT Planacon XP85012 from Photonis \cite{XP85012_DataSheet} with sapphire window. 
This is a PMT with a bialkali photocathode, \unit{25}{\micro\meter} micro-channel diameter, 
active area of \unit{53\times53}{\mm\squared} and $8\times8$ anode pad structure, resulting in 
a pad size of \unit{5.9\times5.9}{\mm\squared} and a pitch of \unit{6.5}{\mm}. 
To operate the PMT, we use the high-voltage divider similar to the recommended one, but with resistance values  providing higher 
voltage between photocathode and MCP: 
$R1=\unit{2.2}{M\ohm}$, 
$R2=\unit{10}{M\ohm}$, and 
$R3=\unit{1}{M\ohm}$, 
see Fig.~\ref{fig:divider}.
We operate the PMT with a high voltage up to \unit{2}{kV}, which corresponds to a PMT amplification of 1~--~2~$\cdot10^6$
 depending on the PMT.

\subsection{Readout Electronics}
\label{sec:readout}
\begin{figure}
\includegraphics[width=.42\textwidth]{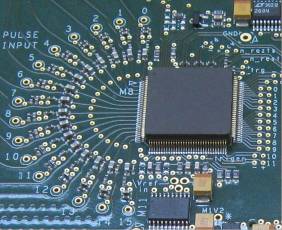}
\hfill
\includegraphics[width=.56\textwidth]{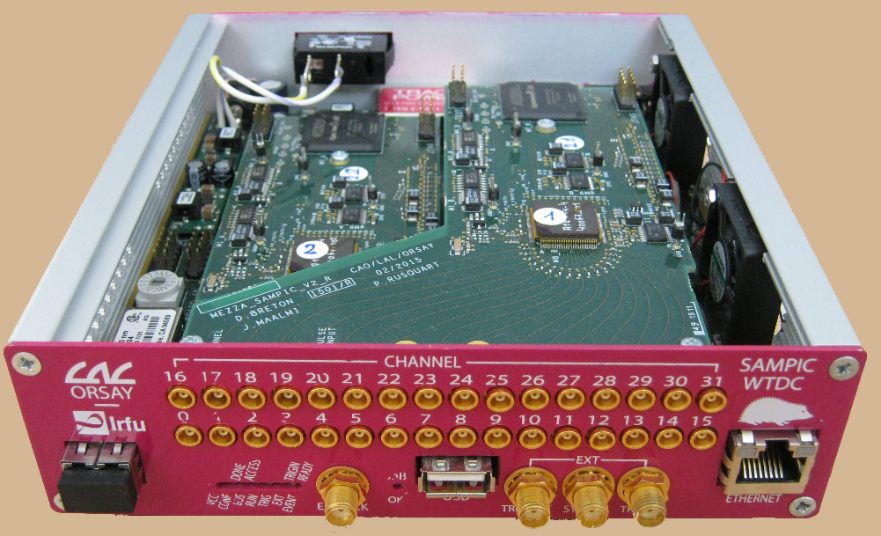}
\caption{\label{fig:sampic}
SAMPIC\_V3C digitization chip (left) and SAMPIC 32-channel module (right).}
\end{figure}

\begin{figure}
  \begin{minipage}[t]{.42\textwidth}
    \centerline{\includegraphics[width=\textwidth]{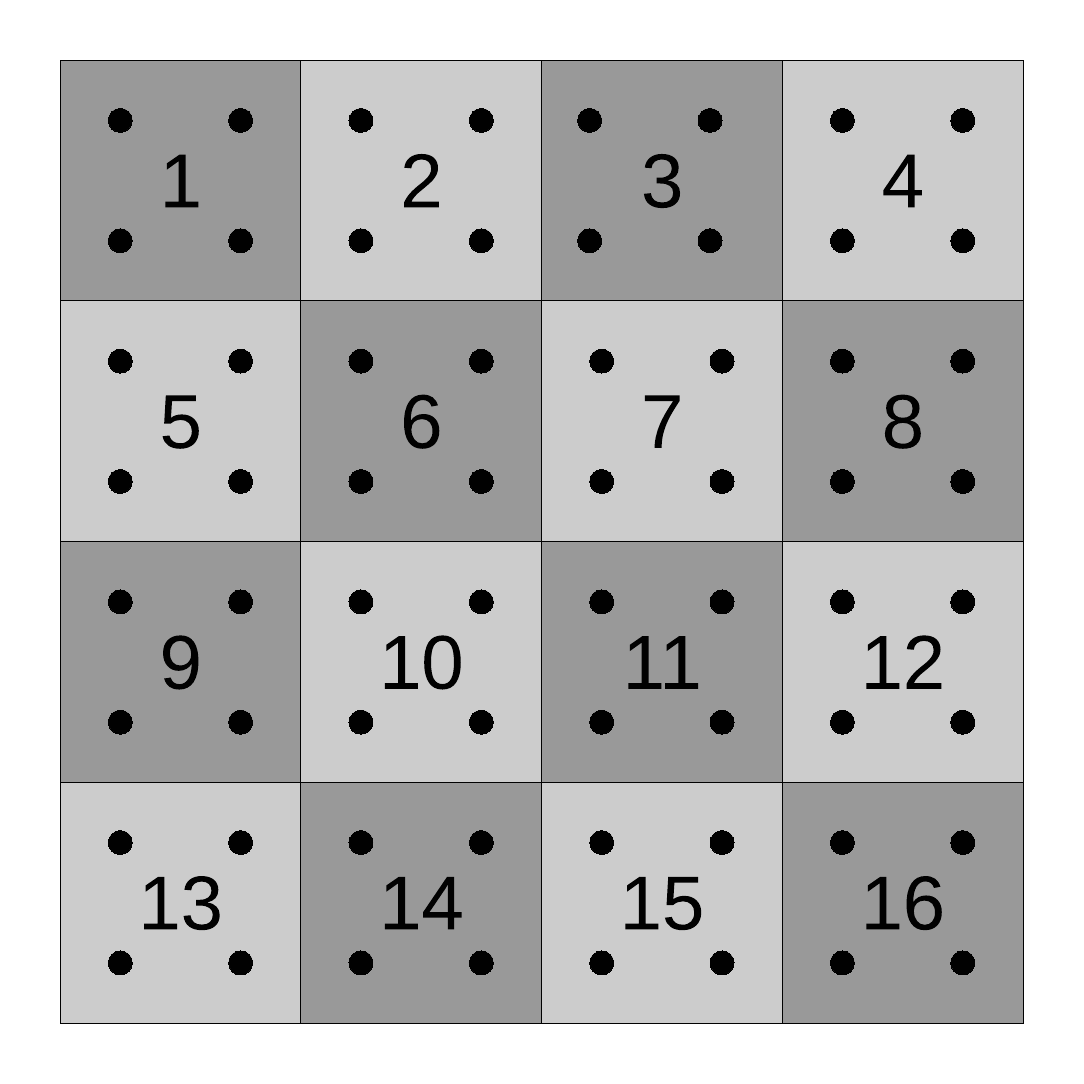}}
    \caption{Readout channels numbering. Each point represents one of the 64 PMT anodes. Four anodes 
      are connected to each readout channel.
      \label{fig:anodes_structure}}
  \end{minipage}
  \hfill
  \begin{minipage}[t]{.56\textwidth}
    \centerline{\includegraphics[width=\textwidth]{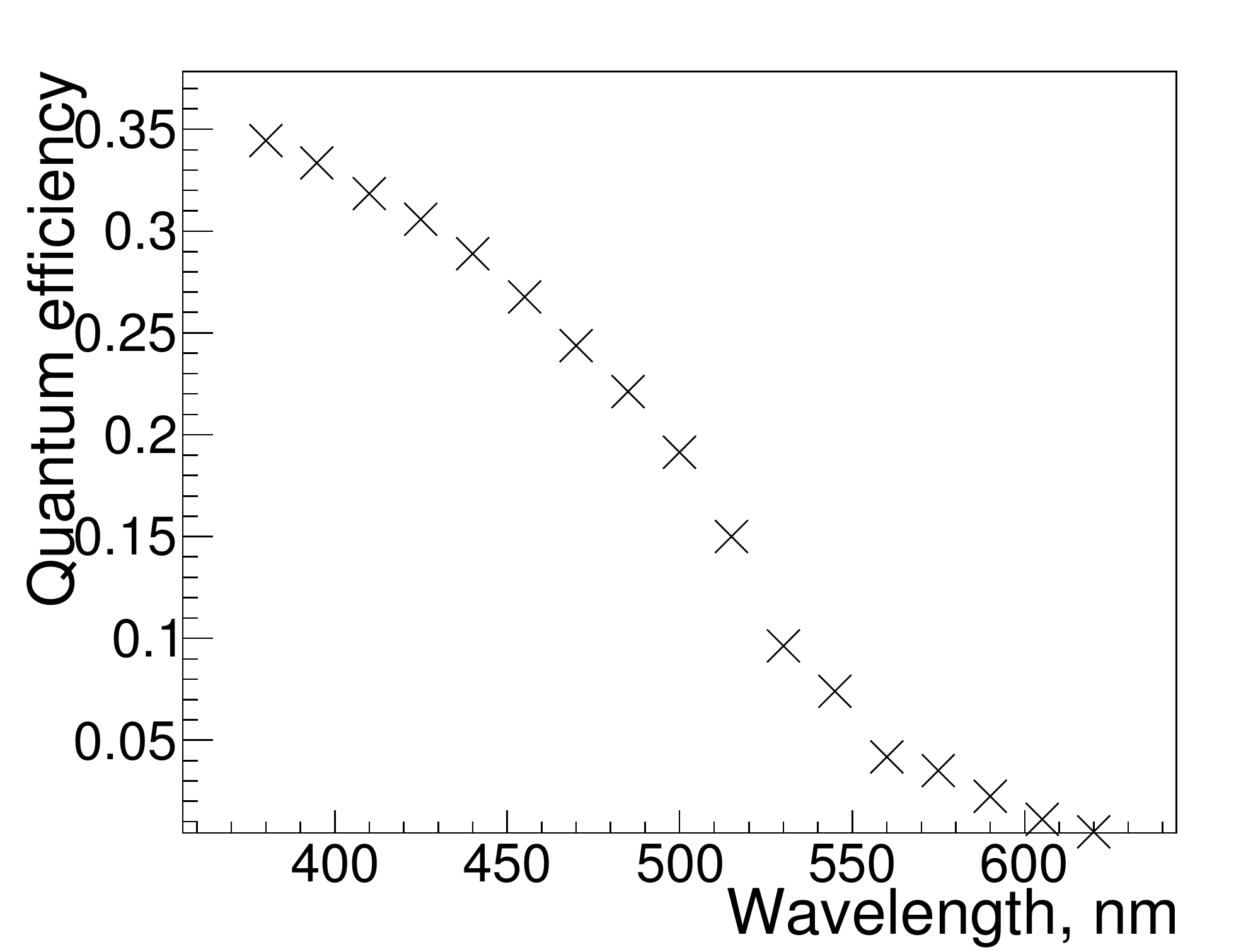}}
    \caption{\label{fig:QE_simu}
      Photocathode efficiency used in Geant4 simulation.}
  \end{minipage}
\end{figure}

With an amplification of about $10^6$ the  MCP-PMT provides a single photoelectron signal with an amplitude of several \milli\volt.
The digitization module SAMPIC, Fig.~\ref{fig:sampic}, 
requires to have a signal in the range of 50 -- 1000 \milli\volt, such that an additional amplification 
of PMT signals is necessary. For the  Planacon XP85012 we measure a signal rise time 
of \unit{0.65}{\ns}~(10\% -- 90\%) 
 and a typical signal duration of about \unit{2}{\ns} (FWHM\footnote{Full Width at Half Maximum}). 
In order to preserve the signal shape, an amplifier with a bandwidth of at least \unit{700}{MHz} is necessary. 
For the historical reason, we used two types of the commercial amplifiers with 50~\ohm~input impedance: 
ZKL-1R5  with amplification \unit{40}{\db} and bandwidth \unit{1.5}{GHz}, and 
ZKL-2R7+  with amplification \unit{24}{\db} and bandwidth 2.7~GHz.
In the described test we were limited to sixteen digitization channels per PMT, such that we developed a customized printed circuit board (PCB),
that connects four anodes to one readout channel, as it shown in Fig.~\ref{fig:anodes_structure}.

We used one or two detection modules and  readout them with the 32-channel SAMPIC module \cite{Delagnes:2015oda,Breton2016}.
This SAMPIC module contains two 16-channel SAMPIC\_V3C chips, which are based on the concept of Waveform and Time to Digital Converter. 
Each channel of the chip
includes a DLL-based TDC providing a raw time associated with an ultra-fast analog memory sampling
of the signals used for waveform recording and precise timing measurement (as good as a few ps RMS).
 Every channel also integrates a discriminator that can trigger it independently or participate to a more complex trigger,
such as ``OR'' or coincidence between programmable channels.
 A first trigger level is implemented on-chip while a second trigger level (L2) can be performed at the module level (32 channels).
The SAMPIC module provides several sampling frequencies ranging from 1.6 to 8.5 \giga S\per\second.
It is controlled and readout via an associated data-acquisition software which is used to configure SAMPIC module, 
start and stop acquisitions, store recorded data on disk in binary or ASCII format and visualize signal waveform and parameter distributions.
The SAMPIC module transfers raw signal waveforms to the computer, where all necessary calibrations are applied on-the-fly by the software.
In addition, the software performs one of the three hit time extraction algorithms using the waveform data: fixed threshold, 
constant fraction discriminator (CFD) or multiple CFD.
In this work we run at 6.4 GS/s and use the CFD algorithm, as implemented in the SAMPIC software with a 0.5 amplitude fraction.
The data to store includes signal waveforms, calculated time and amplitude, and selected SAMPIC parameters. 
To reduce the required disk space and increase the i/o data rate, the signal waveforms can be omitted in the file recording.

\section{Efficiency Measurement} 
\label{sec:eff}
In this work we are studying the feasibility of a PET detection module based on the detection of Cherenkov radiation. Reaching a high efficiency allows to use less annihilation events to obtain good quality image and hence to inject smaller quantity of the radiopharmaceutical and therefore reduces the dose delivered to the patient.

As discuss above (sec.~\ref{sec:cherenkov_rad}), the \PbF\ crystal with thickness 10 mm provides a 75\% efficiency to convert 511~keV\ photon. 
Increased thickness provides higher efficiency, but degrades the time resolution. The photon collection in the crystal depends to a large extend on the quality of the optical interface. The refraction index of the crystal ranged from 1.94 at 300 nm to 1.76 at 600 nm (see e.g. \cite{Anderson:1989uj}). The optical gel OCF452, used as an optical medium, is transparent for the photons with a wavelength larger than 300~nm and has a refractive index of 1.55 at 400 nm \cite{OpticalGel}. Due to the significant mismatch between crystal and gel refractive indexes,
photons with incident angles more than critical angle $\theta_c = \arcsin{(n_{Gel}/n_{\PbF})} \sim \unit{53}{\degree}$ at 300~nm are reflected (total internal reflection). This reduces the transition probability through the optical interface to the photocathode. To improve the photon collection efficiency it would be better to use an optical medium with higher refraction index. Unfortunately, a general trend observed for optical media is that larger refraction index corresponds to a larger cutoff wavelength. 
The Cherenkov radiation peaks at blue and ultraviolet values and hence, higher cutoff leads to significant losses of photons with short wavelength.
We evaluated the alternative optical ``Meltmount Media'' with refraction index 1.73 at 400~nm~\cite{Meltmount} and which is transparent for $\lambda>\unit{400}{\nano\meter}$.
The Geant4 simulation shows the same efficiency of 23.1\% as for the OCF452 gel, so the improvement in total internal reflection is counterbalanced by the reduced spectral transparency. 

The next step in the detection is the conversion of the optical photon
to electron(s) in the PMT photocathode. Planacon XP85012 contains a
Bialkali photocathode. According to the datasheet~\cite{XP85012_DataSheet} the highest photocathode quantum
efficiency (QE) is of the order of 22\% at 380~nm (the Fresnel
reflection at the window boundary is taken into account). In the
Geant4 simulation we implement the photocathode as a thin metallic
layer with the refraction index corresponding to the measurements
in~\cite{Motta2005a}. The simulation accounts naturally for the
reflection at the windows boundary and for the total internal and
Fresnel reflection from the photocathode surface due to its high
refraction index (2.7 at 440~nm). We implemented the photocathode
quantum efficiency using the datasheet data~\cite{XP85012_DataSheet}
corrected for the Fresnel reflection both at the interface air-window
and window-photocathode, see Fig.~\ref{fig:QE_simu}. Since the details
of the photocathode deposition on the sapphire window are not known,
the simulation is not expected to reproduce the absolute efficiency
with high precision. In addition, the price to pay, when using a
MCP-PMT is an extra collection inefficiency for electrons emitted from
the photocathode. In the simulation, we assumed a collection
efficiency of 60\%, typical for these MCP-PMTs, see for example
\cite{Hirose2015Jul,Barnyakov2017}. This value depends on the
open-area of the PMT (ratio of channels area to total MCP area) and
from the probability of the electron to backscatter from the top of
MCP and be collected after that in a pore. In particular, the
efficiency to collect the backscattered electrons is limited by the
time delay of generated signals, due to the finite duration of the
coincidence time window used in the measurement, see
section~\ref{sec:pmt_reso}.

\begin{figure}[!t]
\centerline{\includegraphics[width=.9\textwidth]{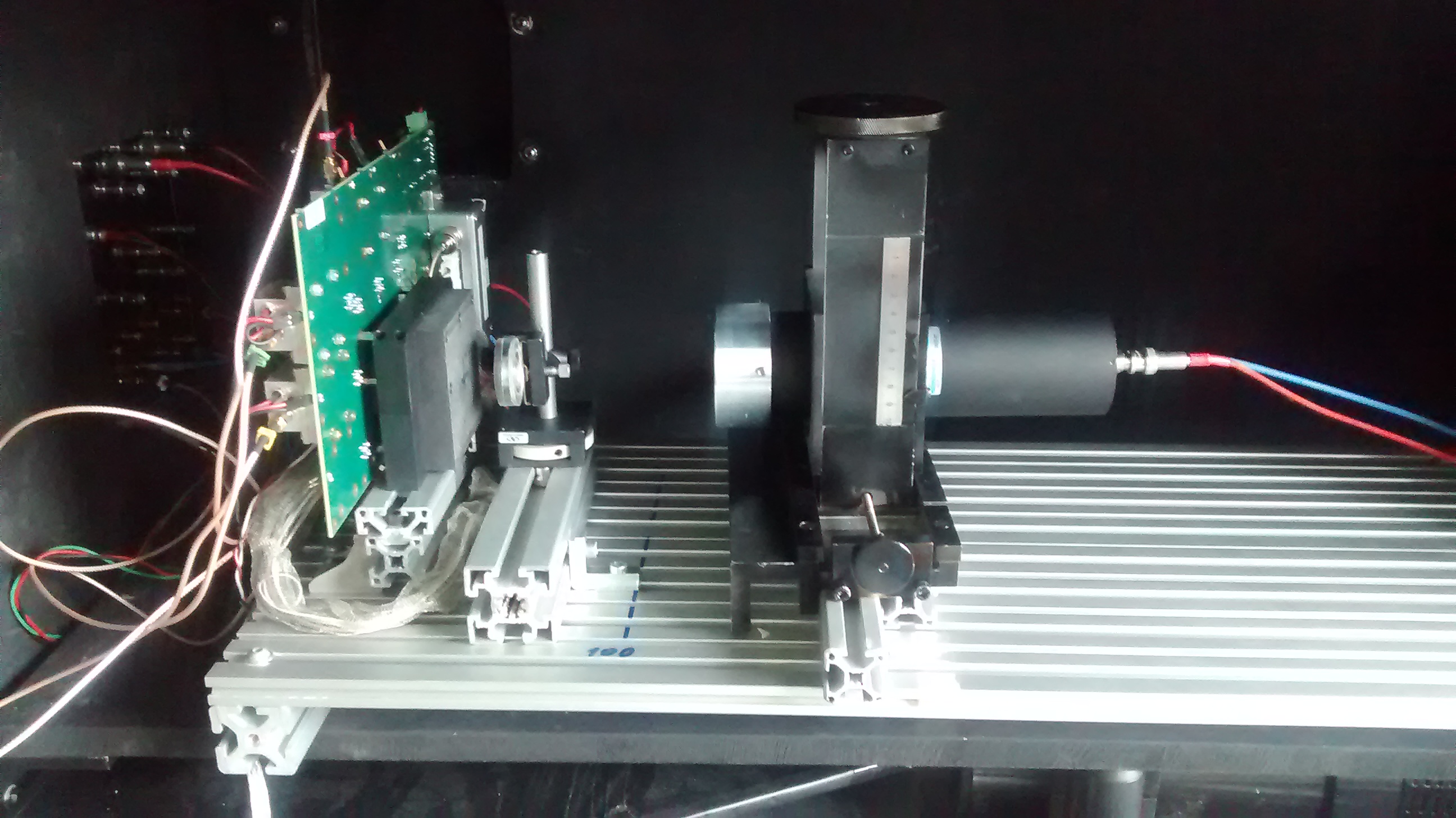}}
\caption{Setup for the detection efficiency measurement. 
From left to right : the \PbF\ detection module (black squared block) with amplifiers PCB (green plate);
the \Na~radioactive source (transparent plastic disk); YAP spectrometer (black cylinder with the metallic ring).
\label{fig:setup_eff} }
\end{figure}

\begin{figure}[ht!]
\centerline{\includegraphics[width=\textwidth]{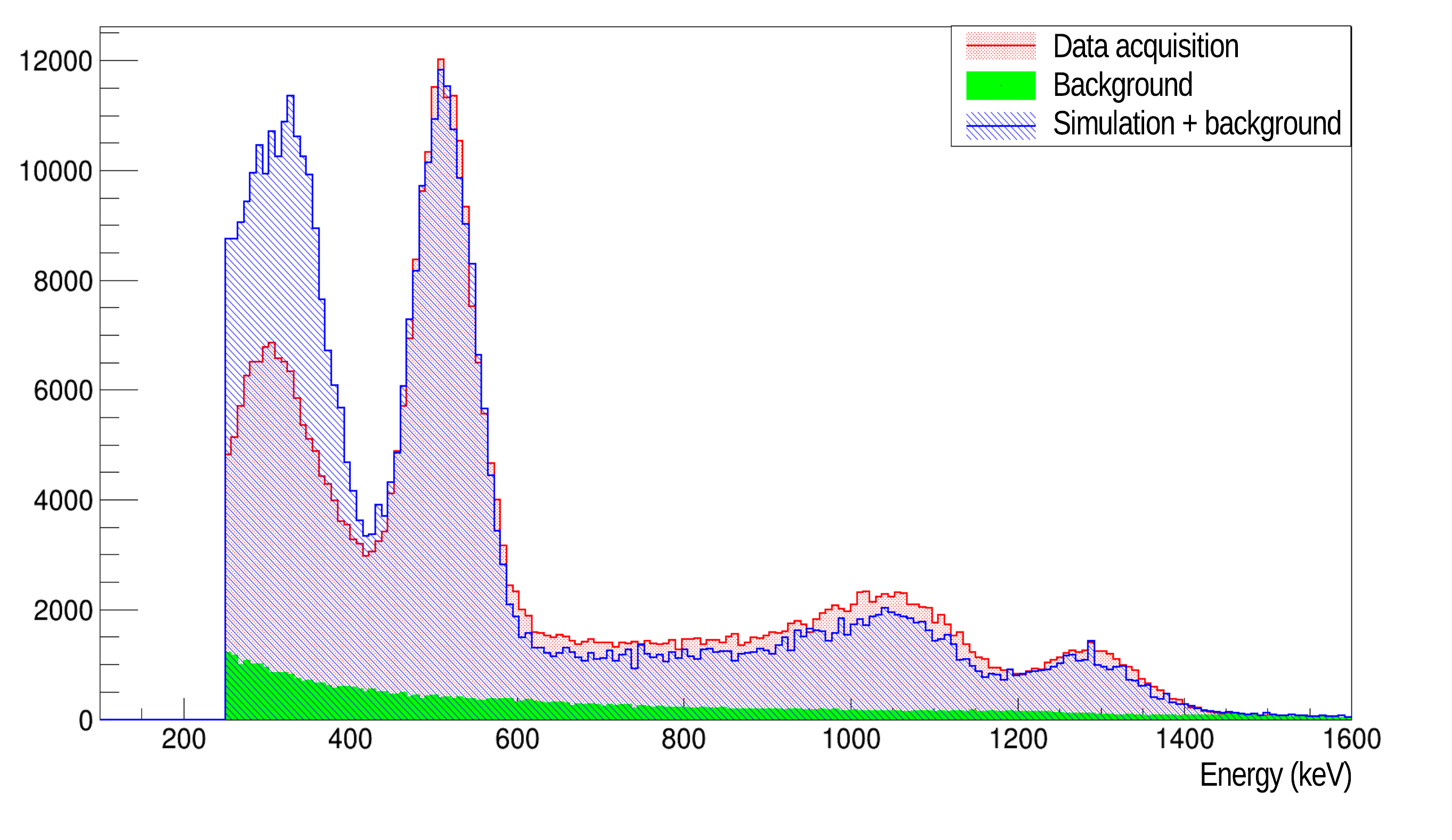}}
\caption{Measured and simulated spectra, produced by a \Na~radioactive source in the YAP spectrometer. 
The pink histogram is the measured spectrum. The green histogram corresponds to the data collected without any radioactive source. 
The blue histograms is the energy distribution in simulation added to the spectrum without radioactive source.
\label{fig:yap_energy}
}
\end{figure}

We measure the detection efficiency with the ``tag-and-probe'' method.
For this we use a \Na~radioactive source which emits a positron simultaneously with a 1.27~MeV photon.
The positron annihilates with an electron in the encapsulating plastic and produces two 511~keV back-to-back photons. 
The first 511~keV photon is detected by the ``tag'' detector, a gamma spectrometer with YAP:Ce crystal.
The second 511~keV photon is detected (or not) by the ``probe'' detector, i.e.  by the \PbF\ detection module,
see Fig.~\ref{fig:setup_eff}. 
We calculate the detection efficiency $\varepsilon$ in the  \PbF\ detection module as:
\begin{equation}
  \varepsilon= \dfrac{N_{\PbF}}{N_{\rm YAP}} ,
  \label{eq:eff}
\end{equation}
where $N_{\rm YAP}$  is the number of events with 511~keV photon conversion in the YAP spectrometer, 
and $N_{\PbF}$ is number of events in \PbF\ detector registered in coincidence with the YAP spectrometer.
The distribution of energy deposition measured by the YAP spectrometer is shown in Fig.~\ref{fig:yap_energy}.
Only events from range $[\unit{511}{keV}-\Delta\ , \unit{511}{keV} + \Delta]$ are accounted in $N_{\rm YAP}$, where
$\Delta$ denotes the half-width of the energy range selection around the photoionization peak, typically 40~keV.

Events, recorded with the YAP spectrometer, contain not only signals from 511 keV photon conversion, 
but also contributions from natural radioactivity and cosmic rays as well as PMT noise. 
These contributions are quantified by recording spectra without any radioactive source. 
An additional background contribution is present due to the Compton scattering of 1.27~MeV photon in the YAP spectrometer or in the environment. 
The corresponding number of events under the 511 keV photon conversion peak is estimated by a simulation where only $1.27$ MeV photon present.
As one can notice in Fig.~\ref{fig:yap_energy}, the experimental
spectrum is not described well 
by a simulation at values higher than 600~keV and especially at low values. 
This is because of the environment simulation, i.e. the description of all supporting elements, 
the metallic test bench, etc.  These elements generate additional scattered photons, which are detected by the module.
While these eventsare mainly outside the energy selection range, a small part of them could affect our selection.
The observed mismatch between the blue and the pink histograms above 600 keV is accounted for by adding a systematic uncertainty of $\pm 1\%$.

As the detector does not measure the photon energy deposit, we cannot distinguish events from 511 keV or 1.27 MeV photons. 
In our test we chose distances to minimize the overlap with 1.27 MeV photon and the remaining contribution is estimated with the Geant4 simulation.
We compute a correction factor of $0.940 \pm 0.005$\ .

\begin{figure}
\includegraphics[width=.48\textwidth]{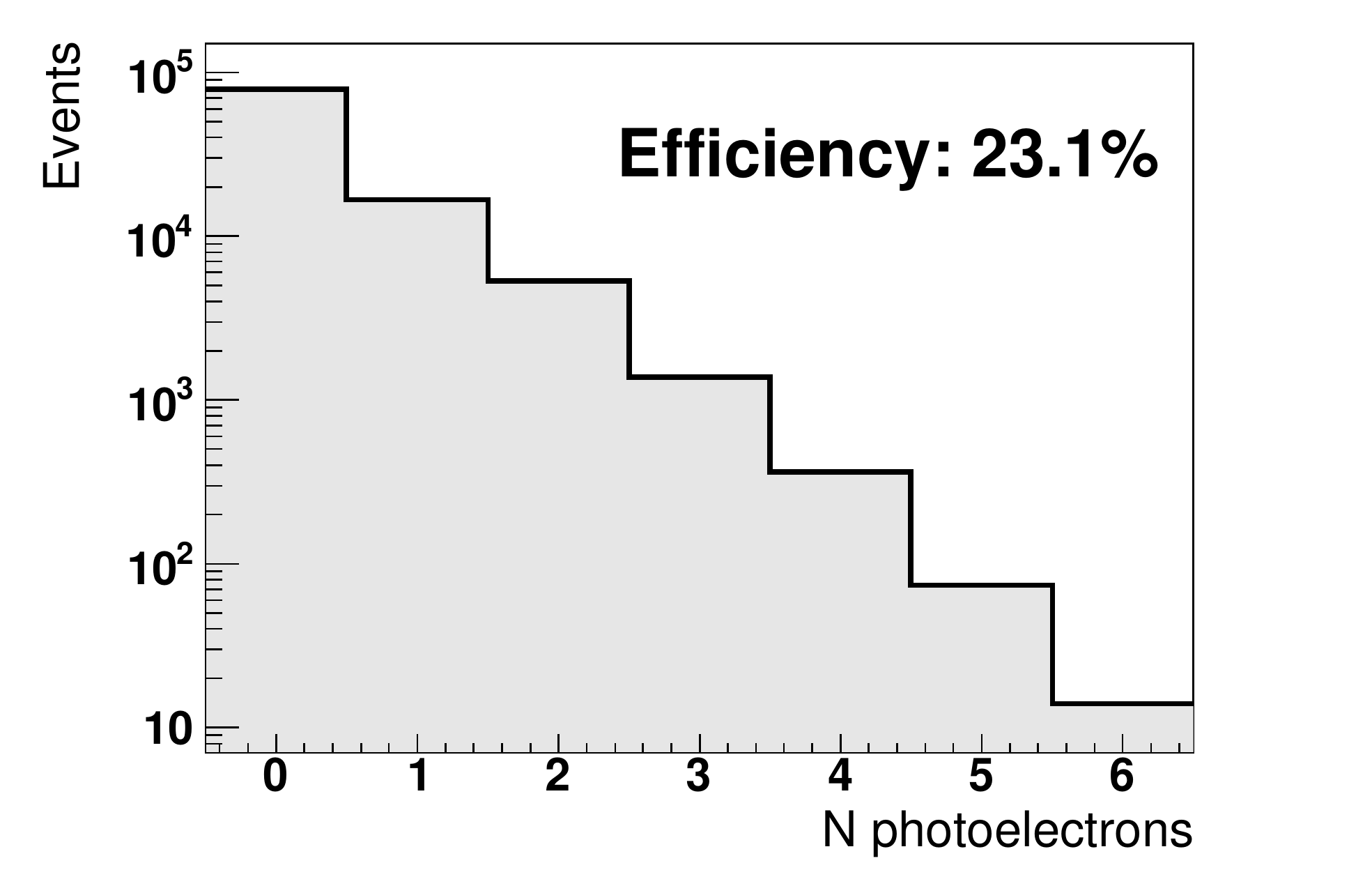}
\hfill
  \includegraphics[width=.48\textwidth]{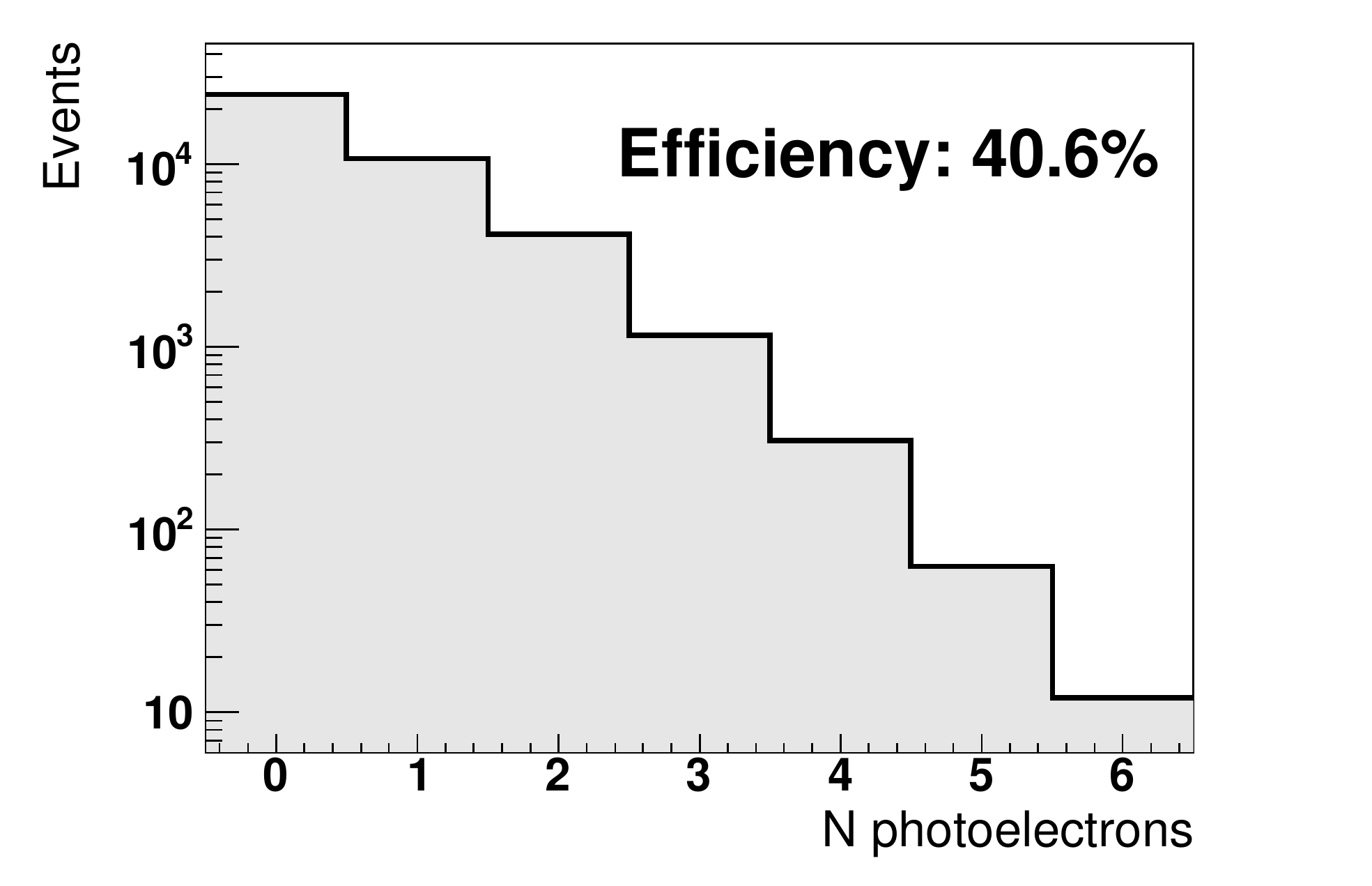}
\caption{Simulated number of photoelectrons produced at the photocathode
(accounting for the photocathode quantum efficiency and photoelectron collection efficiency)  
for all events (left)
  or for events, in which the 511 keV photon was converted through the photoionization (right). 
\label{fig:n_photoe}
}
\end{figure}

Finally, taking into account all corrections mentioned above, 
we compute a global detection efficiency to be:
\[
	 \varepsilon = (23.9 \pm 0.2~(\rm stat) \pm 1.0~(\rm syst))~\%
\]
This number has to be compared with the Geant4 estimation of 23.1$\%$.
As mention above, this estimation has significant uncertainties
related to the description of the photocathode quantum efficiency and
photoelectron collection efficiency.  Nevertheless, the measured
efficiency is in a good agreement with the simulated one.  This number
is much higher than obtained in \cite{Korpar_2013} and compatible with
use of such system for the PET imaging.  This number corresponds to an
efficiency of 30\% to detect an event if the 511~keV photon is
converted in the crystal and of 40\% for events converted through the
photoelectric conversion process, which is of interest for PET
detection.  Fig.~\ref{fig:n_photoe} represents the simulated
distribution of the number of photoelectrons generated at the
photocathode.  As one can see, majority of the detected events have
only one or two photoelectrons and hence it is important to optimize
the optical photon detection in order to reach higher event detection
efficiency.

\section{Time Resolution}
\label{sec:time}
The expected time resolution  for the \PbF\ detection module contains several contributions, 
which in the case of  gaussian distributions,  sum up to a standard deviation (SD) $\sigma$: 
\begin{equation}
	 \sigma^2 = \sigma_{crystal}^2 + \sigma_{PMT}^2 + \sigma_{jitter}^2 +  \sigma_{digit}^2 \ ,
\label{eq:timereso}
\end{equation}
where $\sigma_{\text{crystal}}^2$ is a contribution due to the dispersion of the photon trajectories in the crystal,
$\sigma_{\text{PMT}}^2$ is a contribution due to the transit time spread (TTS) of the PMT,
$\sigma_{\text{jitter}}^2$ is a contribution of the electronics sampling jitter, and
$\sigma_{\text{digit}}^2$  is an electronics  contribution proportional to the signal-to-noise ratio divided by the signal risetime.
In the following sections we describe each of these contributions and estimate the total time resolution by measuring the time difference  
between two identical detection modules.

\subsection{Dispersion of the Photon Trajectories}
\begin{figure}[ht!]
\begin{minipage}[t]{.48\textwidth}
\centerline{\includegraphics[width=\textwidth]{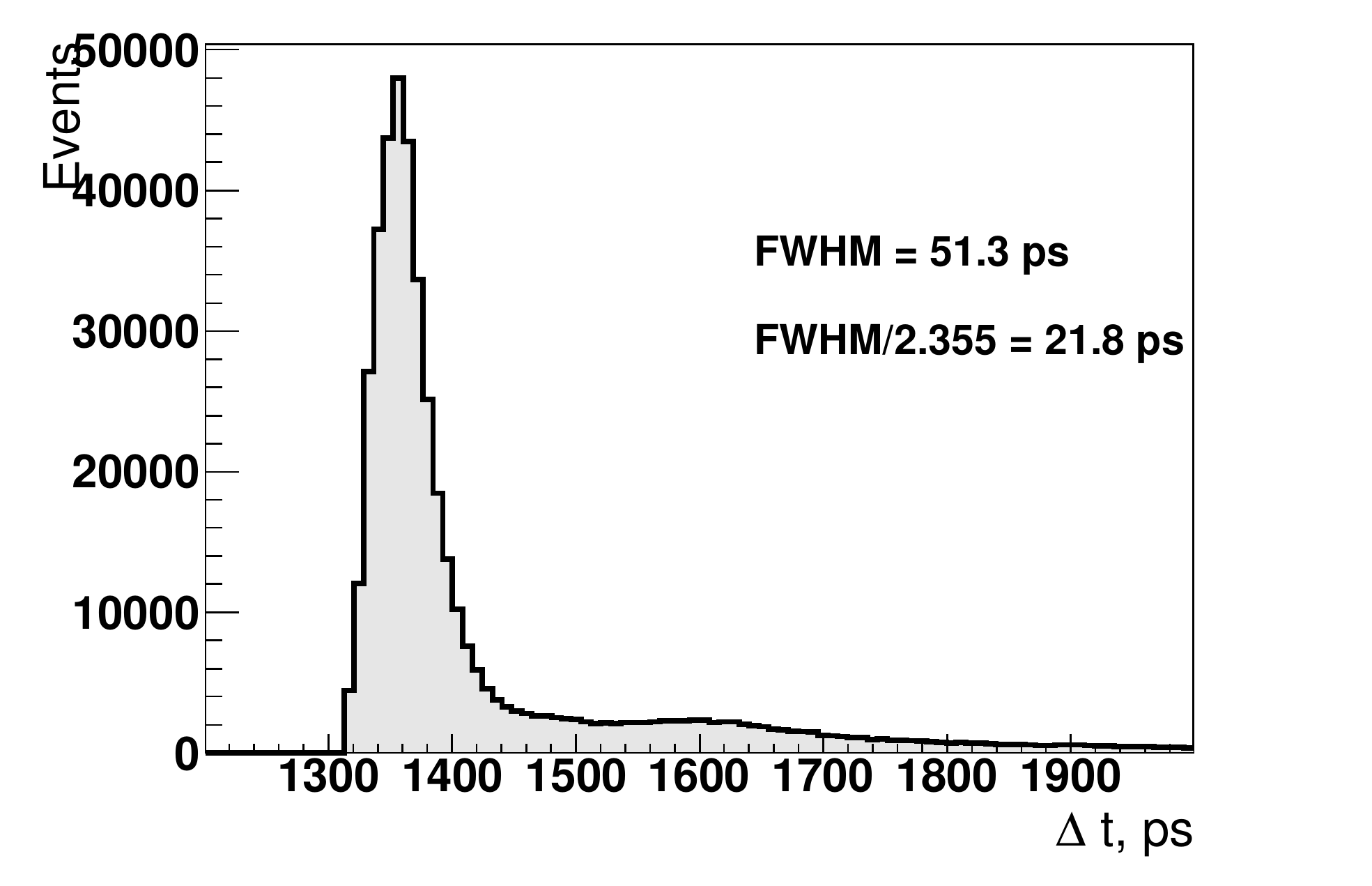}}
\caption{Simulated difference between detection  and  emission time of 511~keV photon.
PMT TTS and electronics time resolution are assumed to be negligible. 
 \label{fig:time1}
}
\end{minipage}
\hfill
\begin{minipage}[t]{.48\textwidth}
\centerline{\includegraphics[width=\textwidth]{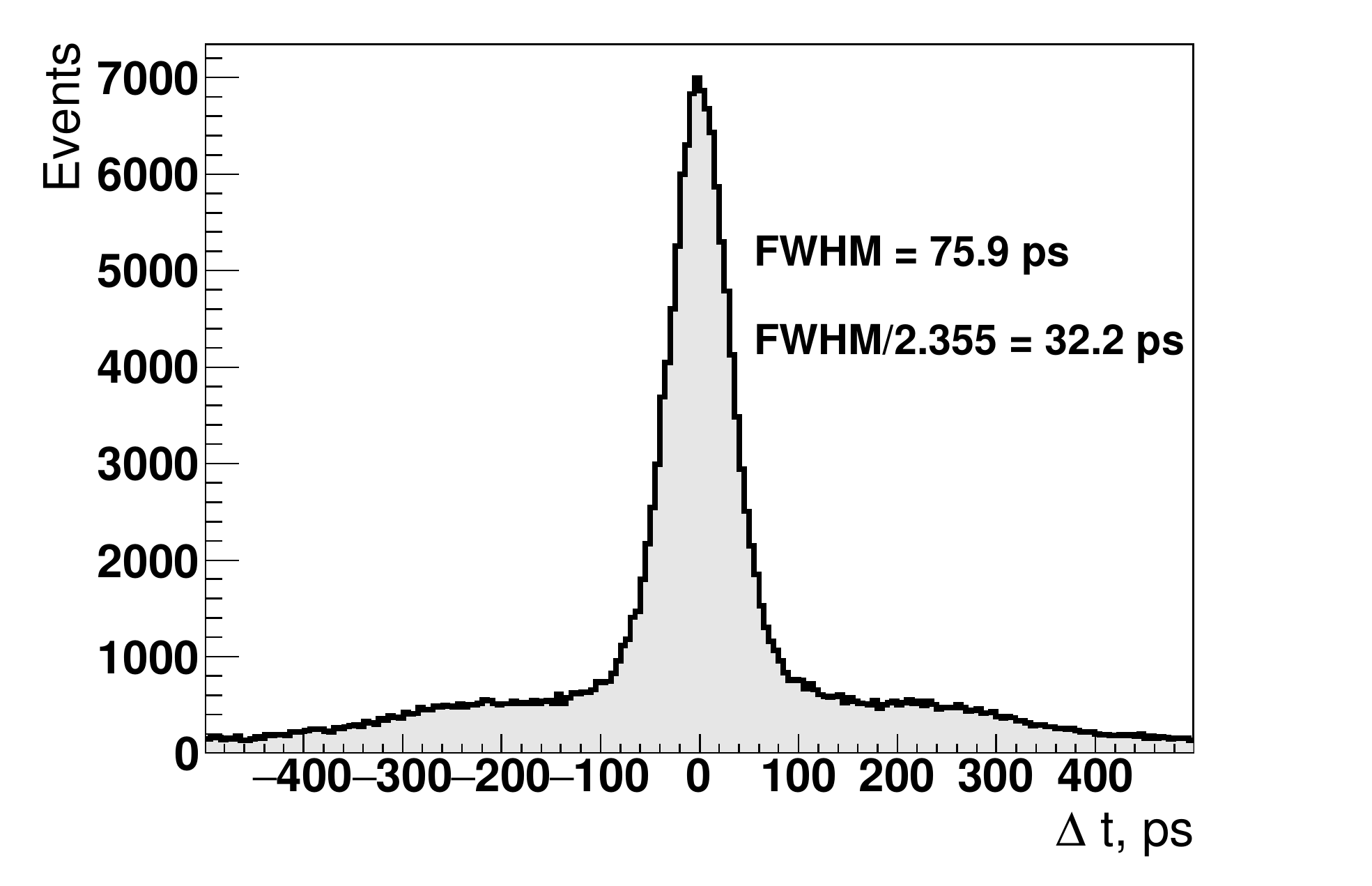}}
\caption{The simulated time difference between detection time of two back-to-back 511~keV photons
in two identical \PbF\ modules.
  PMT TTS and electronics time resolution are assumed to be negligible. 
 \label{fig:twodet_time1}
}
\end{minipage}
\end{figure}

To estimate the contribution from the dispersion of the photon trajectories, 
we use the Geant4 simulation, where we assume negligible 
PMT TTS and no contribution from the digitization electronics. 
The time distribution of the first electron emitted by the photocathode is shown in Fig.~\ref{fig:time1}.
The FWHM is about 50~ps with a tail  due to the Cherenkov photons emitted at large angle, 
leading  to the long travel path with many reflections.
For two back-to-back 511~keV photons emitted simultaneously and detected with two identical modules,
the distribution in the time difference  has CRT of about 76~ps with
30\% of events, which are outside $\pm100$~ps range, Fig.~\ref{fig:twodet_time1} 

\subsection{Readout Electronics Contribution}
We distinguish two contributions related to the digitization of the PMT signal: $\sigma_{\text{jitter}}$ and  $\sigma_{\text{digit}}$.
For the SAMPIC module the sampling jitter is of the order of $\sigma_{\text{jitter}}^2 \sim \unit{3}{ps}$ (SD) \cite{Delagnes:2015oda}.
The $\sigma_{\text{digit}}$ term depends on the noise SD $\sigma_S$, 
which includes contribution from PMT, amplifiers and SAMPIC module and results in a 
noise for each digitization sample of 1.2 mV. The MCP-PMT generates fast signals with a rise time of 0.65~ns. 
For a typical signal amplitude of 100 -- 500~mV the slope of the rising part of a signal $dS/dt$ is about 0.13 -- 0.65 mV/ps,
resulting in $\sigma_{\text{digit}} = \sigma_S/(dS/dt) \le $ 9 -- 2~ps (SD). 
As expected, for signals with small amplitudes, the precision is limited by the contribution $\sigma_{\text{digit}}$, but for the 
signals with high amplitude, it is limited by the constant term  $\sigma_{\text{jitter}}$. 
Overall, we expect the contribution of the read-out electronics to the time resolution to be of the order of
$ \sqrt{\sigma_{\text{jitter}}^2 +  \sigma_{\text{digit}}^2} \sim$ 9.5 -- 3.5~ps (SD),  for the signal amplitude between 100 and 500 mV, similar 
to values reported in \cite{Delagnes:2015oda}.

\begin{figure}[ht!]
\centerline{\includegraphics[width=\textwidth]{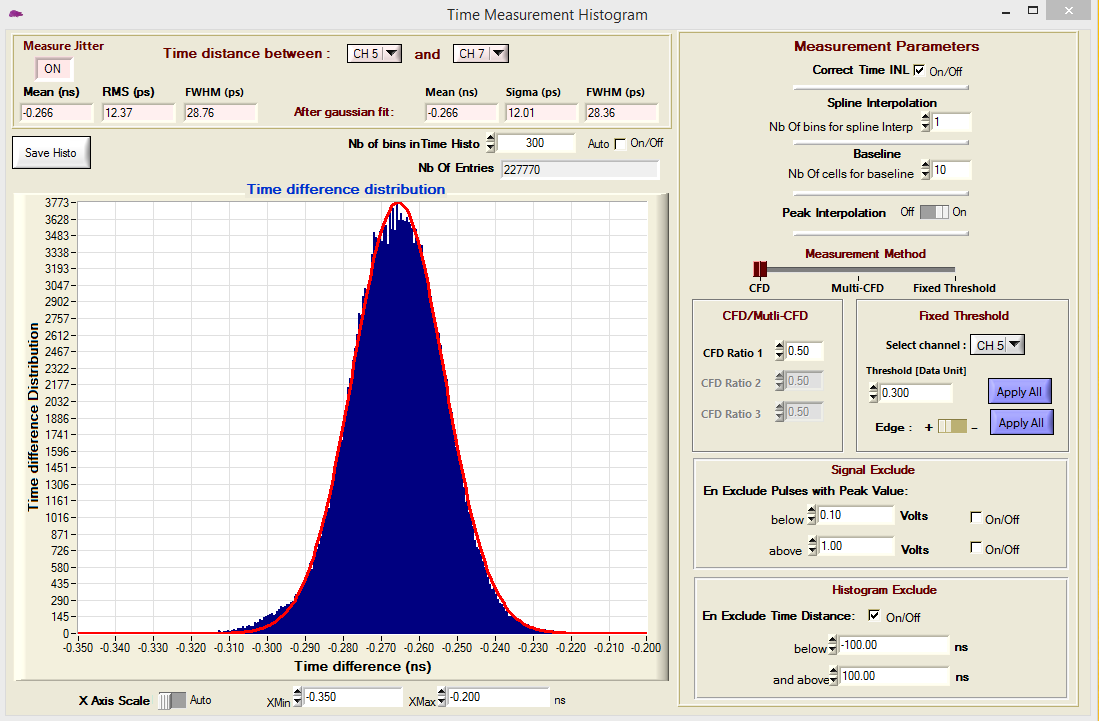}}
\caption{Measured time difference between PMT signal and laser trigger (see text), with the standard deviation of 12~ps,  measured by 
the on-line SAMPIC  data acquisition software. 
 \label{fig:reso_online}
}
\end{figure}

In order to make an independent cross-check of the electronics time resolution, we set-up a dedicated measurement. 
We use a pulsed laser Pilas by A.L.S. as a light source. The laser beam has a Gaussian-like time profile 
with duration of about 20~ps (FWHM) and a jitter of 1.4 ps~\cite{Pilas}. 
We place the laser output at 20~mm distance from the PMT window.
We chose rather high light intensity, of the order of thousand photoelectrons, and operate the PMT at the moderate high voltage of  1400~V.
This results in a stable PMT signal of 150~mV amplitude, amplified to 600 mV in order to match  better the SAMPIC range.
The histogram in time difference between PMT's signal and laser trigger signal 
is recorded by the on-line SAMPIC software, 
see Fig.~\ref{fig:reso_online}. 
The obtained distribution is close to a Gaussian with SD 12~ps. The measured value is larger than expected 
for the signal of 600~mV, but it is obtained with signals from a detector, and possible contribution from the 
PMT TTS.
Additionally, this value contains a contribution from the laser signal jitter, although we expect it to be 
of the order of several picoseconds, due to the large number of photons in the laser beam.

\subsection{PMT Time Resolution}
\label{sec:pmt_reso}
As will be seen in the following, the PMT contribution is the main limiting factor in the overall time resolution. 
To study it in details we build the following system. 
We use the pulsed laser Pilas~\cite{Pilas} collimated by a precise pin-hole 
of 100~\um\ diameter. The distance between laser output and pin-hole is about 100~mm and between pin-hole and PMT window is 10~mm. 
We choose distances and the light intensity in such way that the PMT is working in a single-photon regime with 
a detection efficiency of 3\%. According to the Poisson distribution, 
this corresponds to a ratio of two-photons~/~single-photon events  of 1.5\%.
This number is sufficiently small, that in the following studies
we ignore the presence of events with two photons.

The MCP-PMT is mounted at two-axis X-Z motion system, assembled from two \mbox{X-LRT0100AL-C} linear stages 
from Zaber Technologies Inc. This system allows to move and position the detection  module
with a precision better than 25~\um~\cite{Zaber}.
We realized a detector scan with 1~mm step  and  1.5~s at each position, which leads to the scan duration 
of about two hours per PMT.
For each position we register amplitude and CFD time of PMT signals from anodes and common cathode, 
in coincidence with the laser trigger. We use the so-called ``L2 coincidence'' option of the SAMPIC module, 
with a 20~ns coincidence time window.  
The threshold value for anodes signals is 30~mV.
Parameters of the laser trigger signal (CFT time and amplitude) are also registered by the SAMPIC module.
The data taking rate is between $6\cdot 10^3$ and $2\cdot 10^4$ 
coincidences/s.

\begin{figure}
\subfloat[(x,y)=(12mm, 51mm)] {
\includegraphics[width=.48\textwidth]{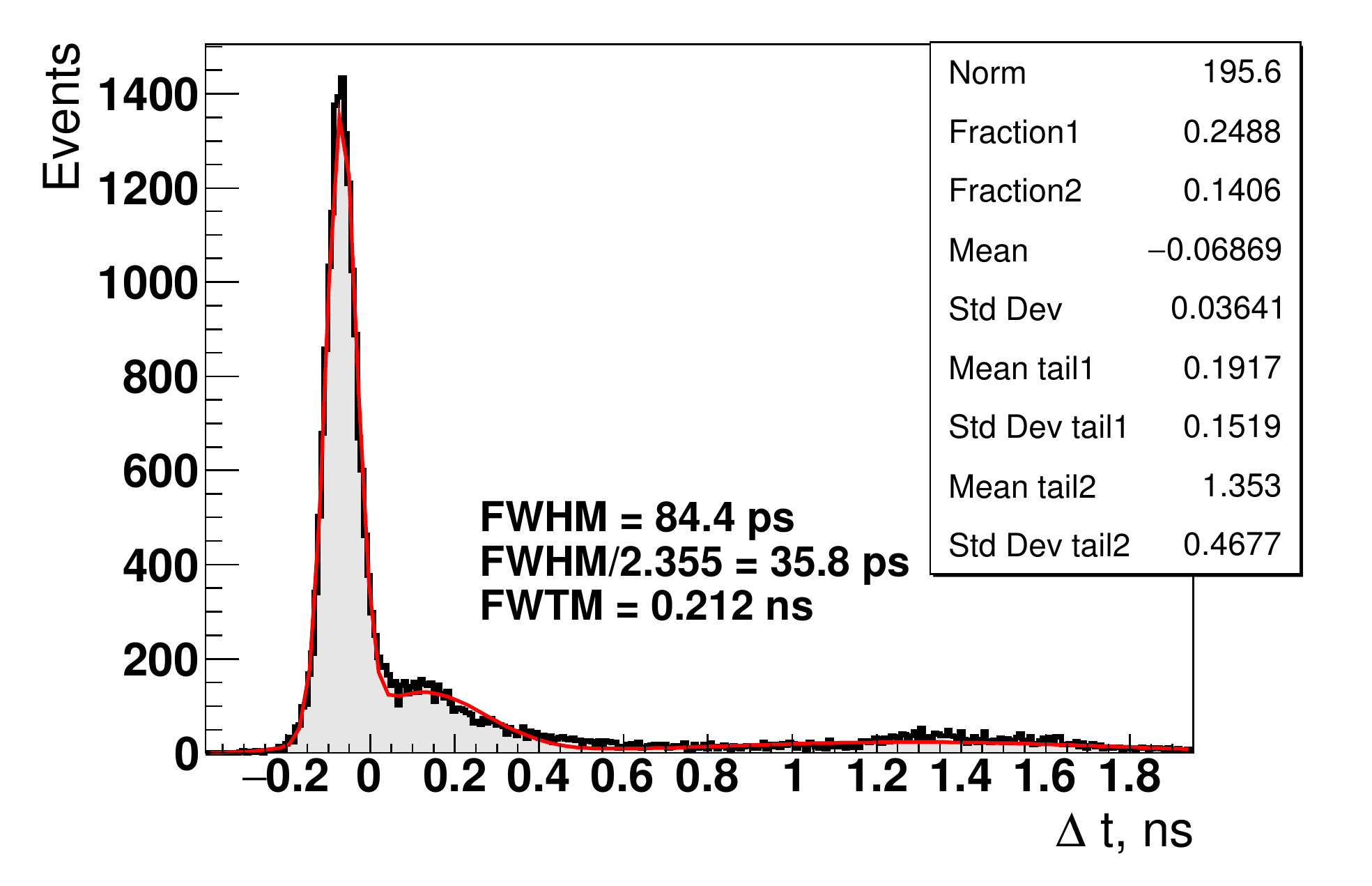}
}
\hfill
\subfloat[(x,y)=(26mm, 40mm)] {
  \includegraphics[width=.48\textwidth]{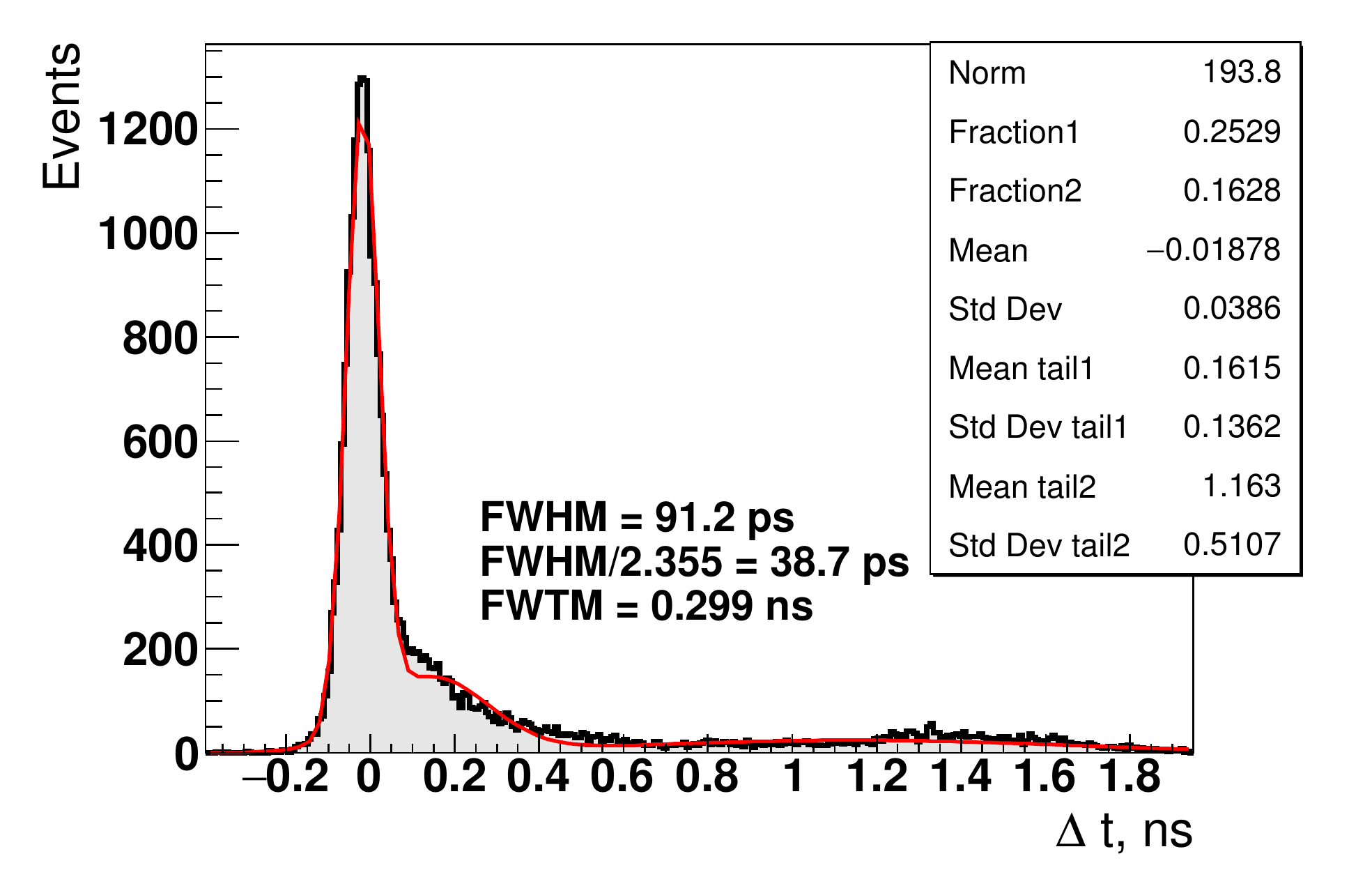}
}
\caption{
Typical difference in time between PMT anode signal and laser trigger measured at two different positions.
The red line represents the triple-gaussian fit function. The FWHM and FWTM (full width at tens of maximum)
are calculated using the histogram and independently from the fit function. 
\label{fig:pixel_dt}
}
\end{figure}

\begin{figure}[ht!]
  \subfloat[Number of channels.] {
    \includegraphics[width=.31\textwidth]{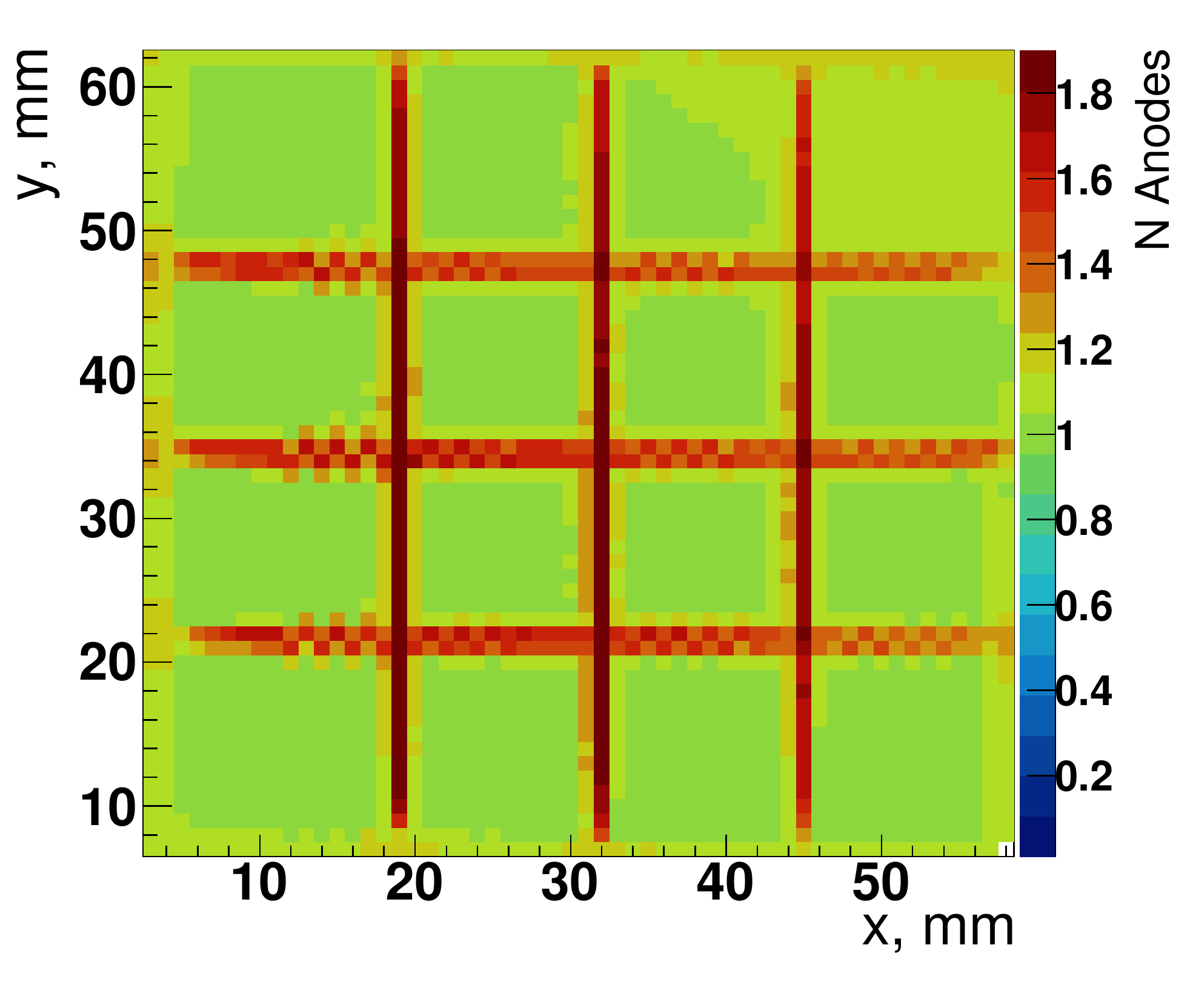}
    \label{fig:nanodes}
  }
  \hfill
  \subfloat[FWHM.] {
    \includegraphics[width=.31\textwidth]{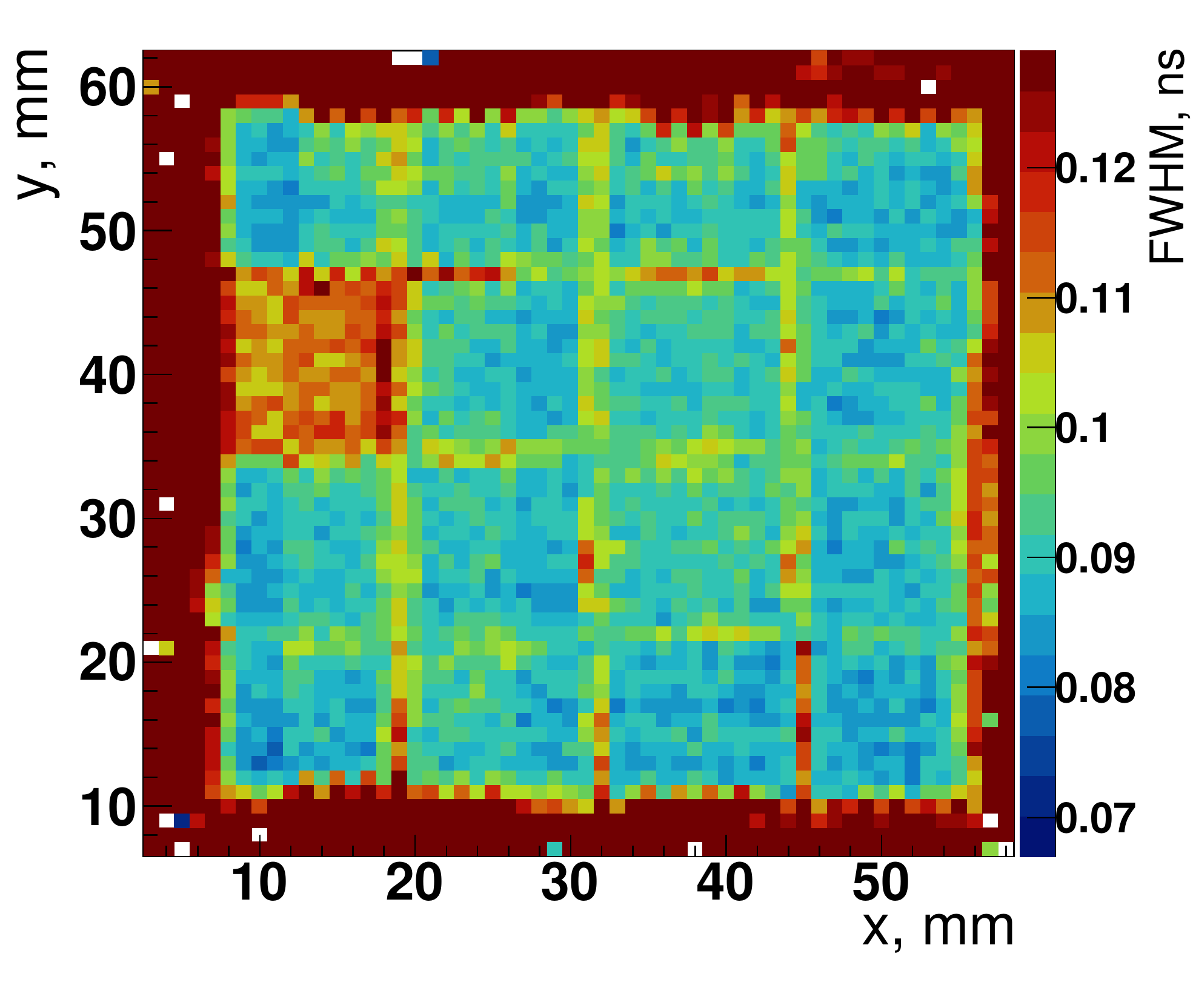}
    \label{fig:fwhm}
  }
  \hfill
  \subfloat[Fraction of events with the time difference $t-t_1$ larger than 100~ps.] {
    \includegraphics[width=.31\textwidth]{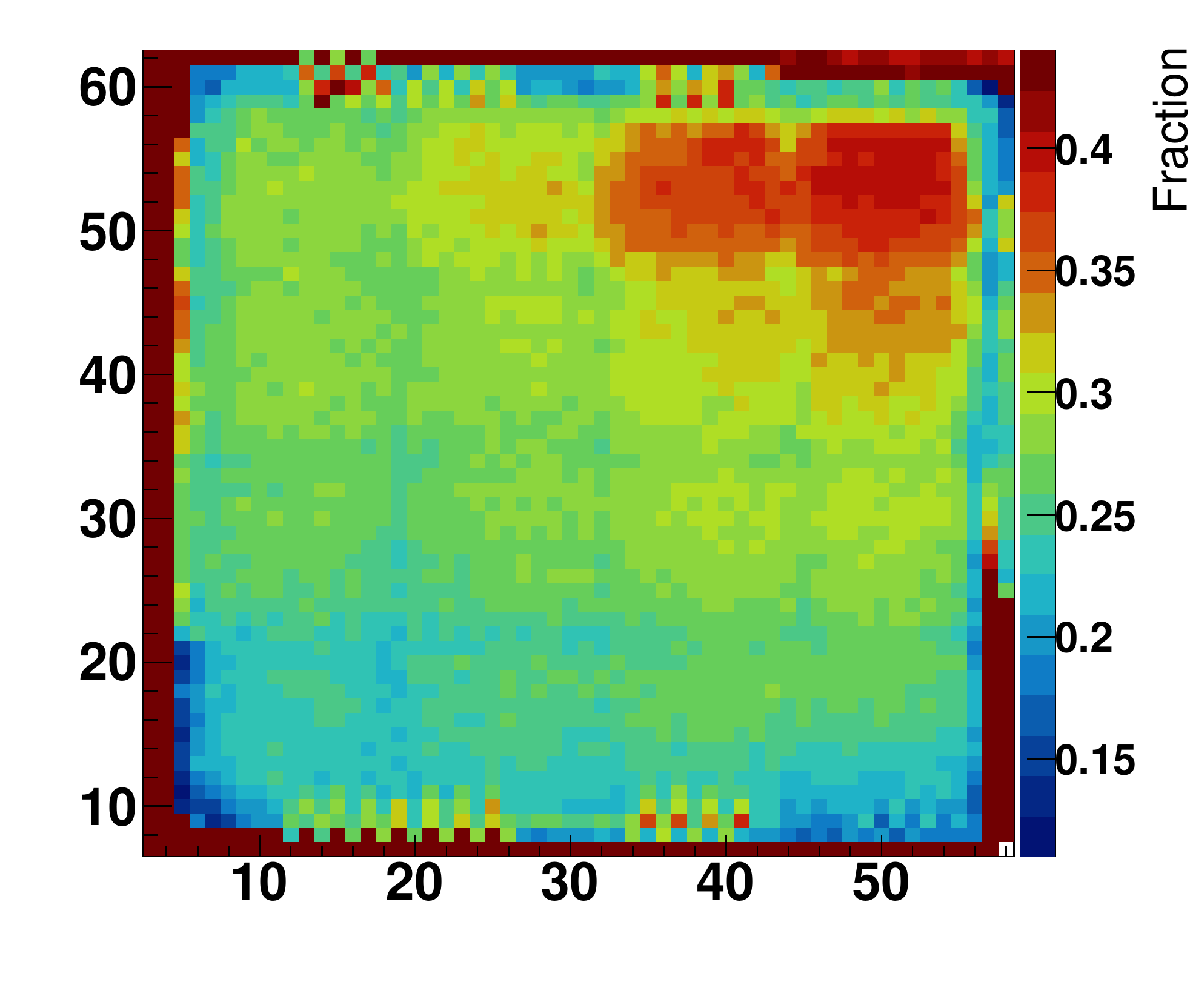}
    \label{fig:fraction}
  }
  \caption{Measured parameters of the time difference distribution  
as a function of the (x,y) coordinates during the laser scan. 
  \label{fig:scan1}
  }
\end{figure}

Typical time difference distributions between anode signals and laser trigger
are shown in Fig.~\ref{fig:pixel_dt}.
The main part of the distribution has the Gaussian-like shape with FWHM between 85 and 100~ps, Fig.~\ref{fig:fwhm}. 
The tail of the distribution is due to the electrons backscattered from the top of the MCP,
see e.g.~\cite{Korpar2008Sep,Lehmann2018}. 
We decided to fit this distribution in the range $[-0.4,1.9]$~ns with the triple-gaussian  function:
\begin{equation}
  f(t) = \frac{n}{\sqrt{2\pi}} \left(
  \frac{(1-f_1-f_2)}{\sigma_1}\ e^{-\frac{(t-t_1)^2}{2\sigma_1^2}} 
  +  \frac{f_1}{\sigma_2}\ e^{-\frac{(t-t_1-t_2)^2}{2\sigma_2^2}} 
  +  \frac{f_2}{\sigma_3}\ e^{-\frac{(t-t_1-t_3)^2}{2\sigma_3^2}} 
  \right) \  ,
\label{eq:reso_func}
\end{equation}
where $n$ is a normalization coefficient, 
$f_1,\ f_2$ are fractions of events in second and third Gaussian terms representing the tail of the distribution,
$t_1$ is the mean of the first term, 
$t_2,\ t_3$ are the additional delays in mean for second and third terms,
$\sigma_1$, $\sigma_2,\ \sigma_3$ are the corresponding standard deviations.
The typical mean value for the first gaussian is $t_1 \simeq -80$~ps, so the backscattered electrons 
with delay up to $t_2 \simeq 2$~ns contribute to the fit.  
As can be seen in Fig.~\ref{fig:pixel_dt}, the function~\ref{eq:reso_func} fit well the main peak, 
but tail is not described perfectly. Nevertheless, we used this function, because it has the advantage to be 
simple and characterizes reasonably well the main features of the distribution.

\begin{figure}[ht!]
  \subfloat[Anode signal] {
    \includegraphics[width=.48\textwidth]{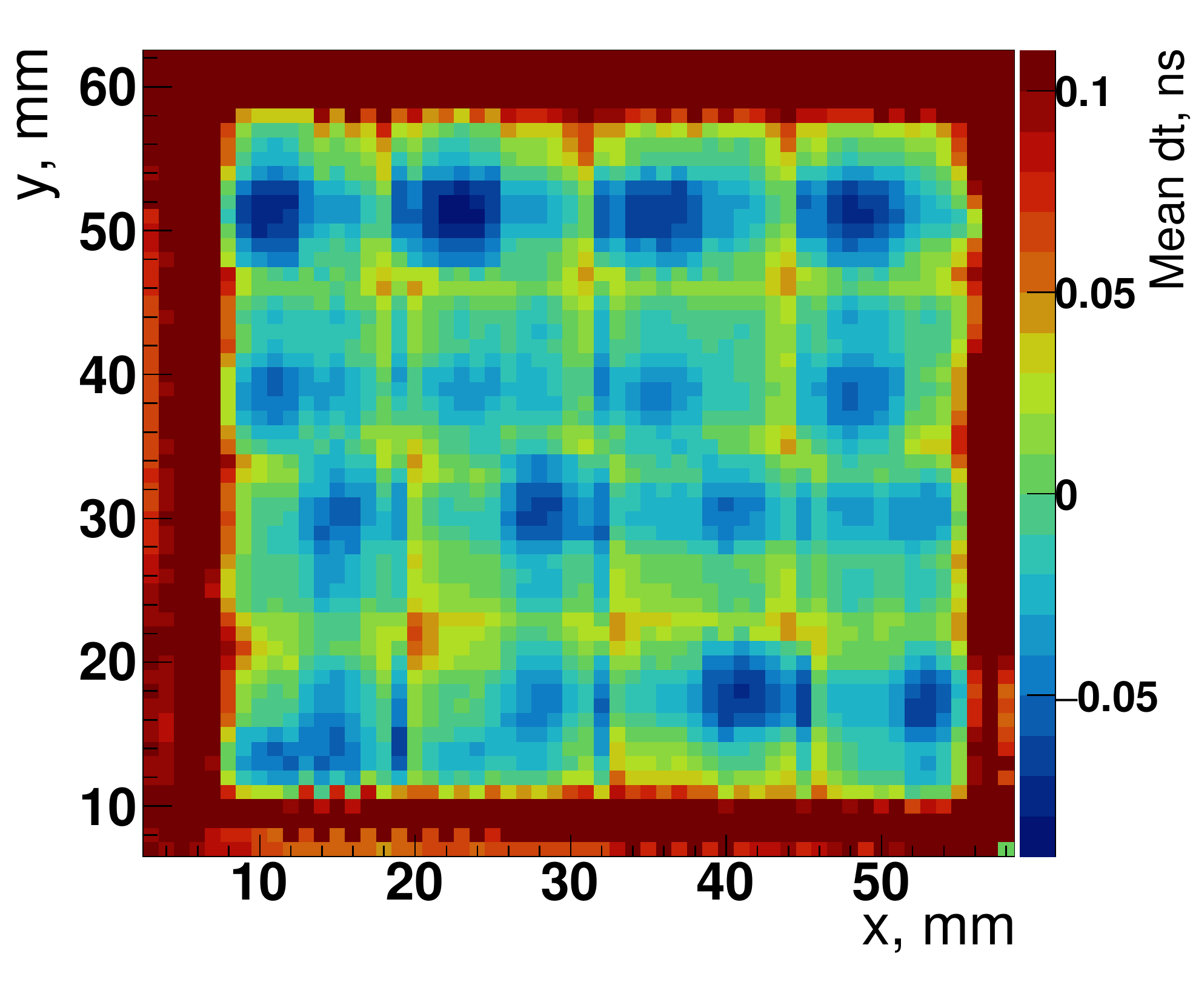}
    \label{fig:mean}
  }
  \hfill
  \subfloat[Cathode signal] {
    \includegraphics[width=.48\textwidth]{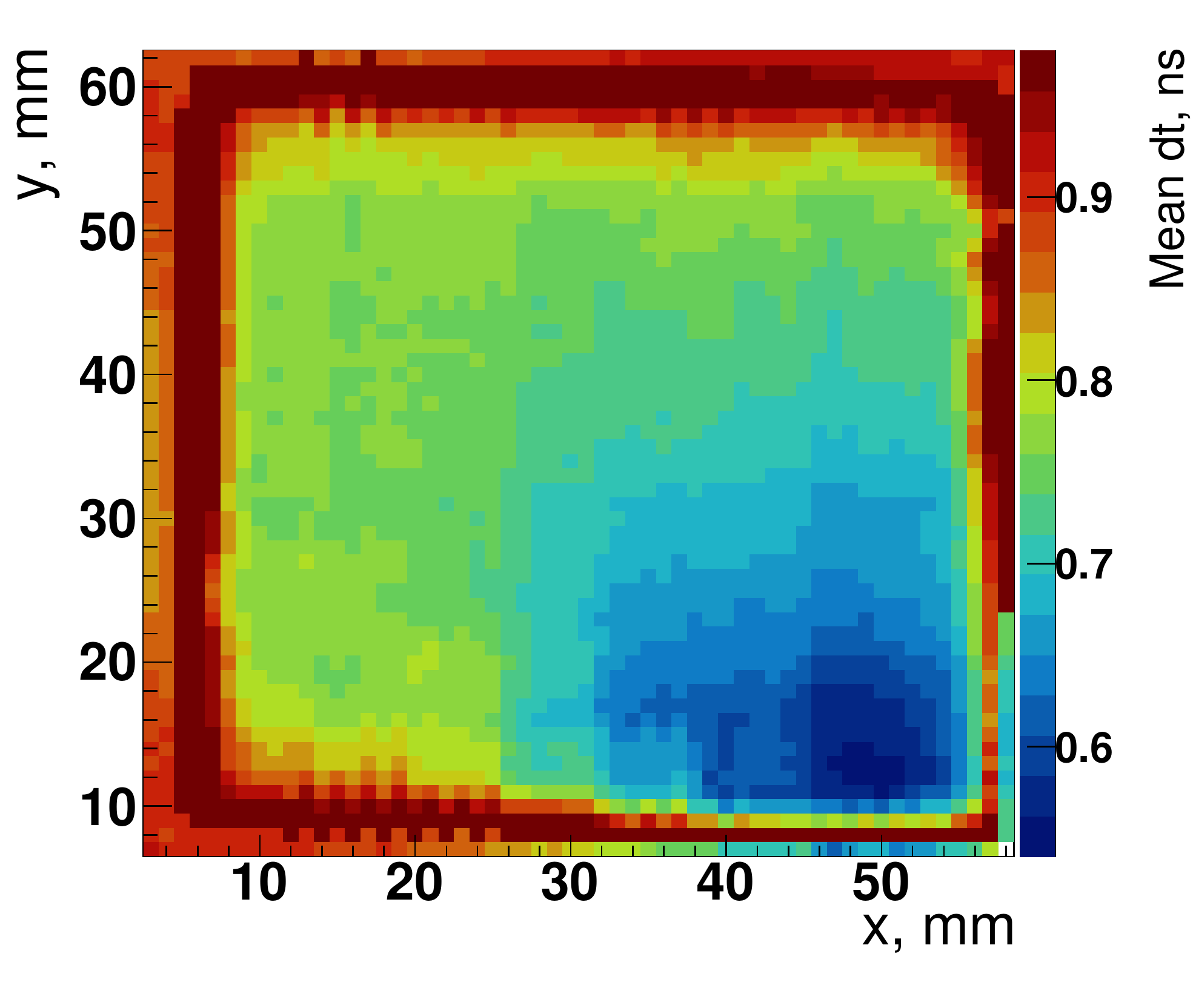}
    \label{fig:mean_cathode}
  }
  \caption{Measured peak timing (parameter $t_1$, in~the~Eq.~(\ref{eq:reso_func})) as a function of the (x,y) coordinates. 
    \label{fig:scan2}
  }
\end{figure}

For each position, the time difference  is fitted using the function~\ref{eq:reso_func} and the 
obtained  parameters are plotted as 2D histograms versus x and y coordinates. 
Several such histograms are presented in Figs.~\ref{fig:scan1} and \ref{fig:scan2} for one of the PMTs.
For example, Fig.~\ref{fig:nanodes} shows the number of channels with signals above 50 mV threshold. 
One clearly identifies regions in the detector where only one channel is triggered at once, 
but at the border between two channels, the two are triggered at the same time (in this case, 
we use the earliest signal time among all triggered as a PMT time). 
This happens due to the so-called charge sharing effect, when the electron shower induces signals simultaneously on two anode electrodes, see e.g. 
\cite{Korpar2008Sep,Li2018Sep,Lehmann2018}.
We observe that charge sharing happens at the distance of 1-2~mm from the pad edge.
The cross-talk due to the capacitive coupling between anodes could cause the similar effects, 
but it affects anode pairs independent of the event position~\cite{Inami2008Jul,Grigoryev2016}. 
For the pixel at the center of anodes, we observe 2.5\% of events with two channels triggered 
simultaneously, so the effect of capacitive coupling is rather small. Additionally, these cross-talk pulses have a differential shape. 
To speed-up the data taking during the scan, we do not register the pulse shape, so we could not apply any shape selection.
In tests with a radioactive source, we register the shape of the signal and
apply additional selection to reduce even more the fraction of cross-talk signals due to the capacitive coupling. 

Fig.~\ref{fig:fwhm} shows that the FWHM of the peak is rather uniform through the entire PMT surface
and has typical value of 85 -- 90~ps. 
At the border between two channels the time resolution is degrading to 100 -- 110~ps due to
the charge sharing effect, but we did not observe any degradation at the border of two anodes when they are connected 
to the same readout channel. This degradation could not be attributed to the signal time walk effect, because we
use the CFD algorithm for time calculation. It rather reflects the more complex interplay between the charge sharing 
mechanism and anode signal formation.
The worse resolution observed for the readout channel with coordinates (x,y)~=~(7--19~mm, 33--46~mm) is attributed 
to an imperfection in the readout PCB design, since we observed it for two different PMTs.
The fraction of events with the time difference $t-t_1$ larger than 100~ps  is around 25\% , but with a degradation in the corner 
of the PMT, Fig.~\ref{fig:fraction}. Such degradation is observed only for one PMT among the two tested.

\begin{figure}
\subfloat[(x,y) = (7--19, 47--60) mm] {
\includegraphics[width=.48\textwidth]{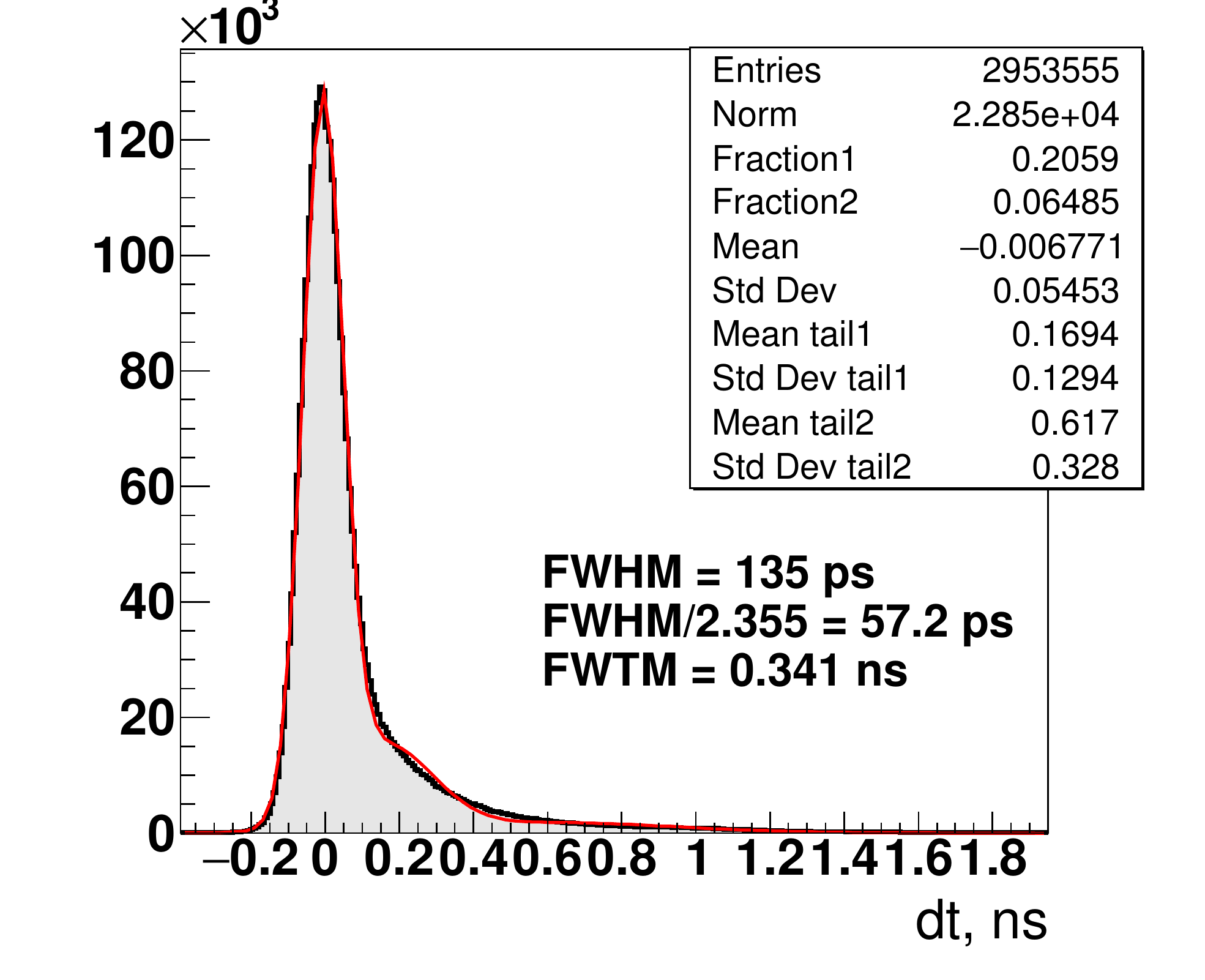}
}
\hfill
\subfloat[(x,y) = (19--32, 35--48) mm] {
  \includegraphics[width=.48\textwidth]{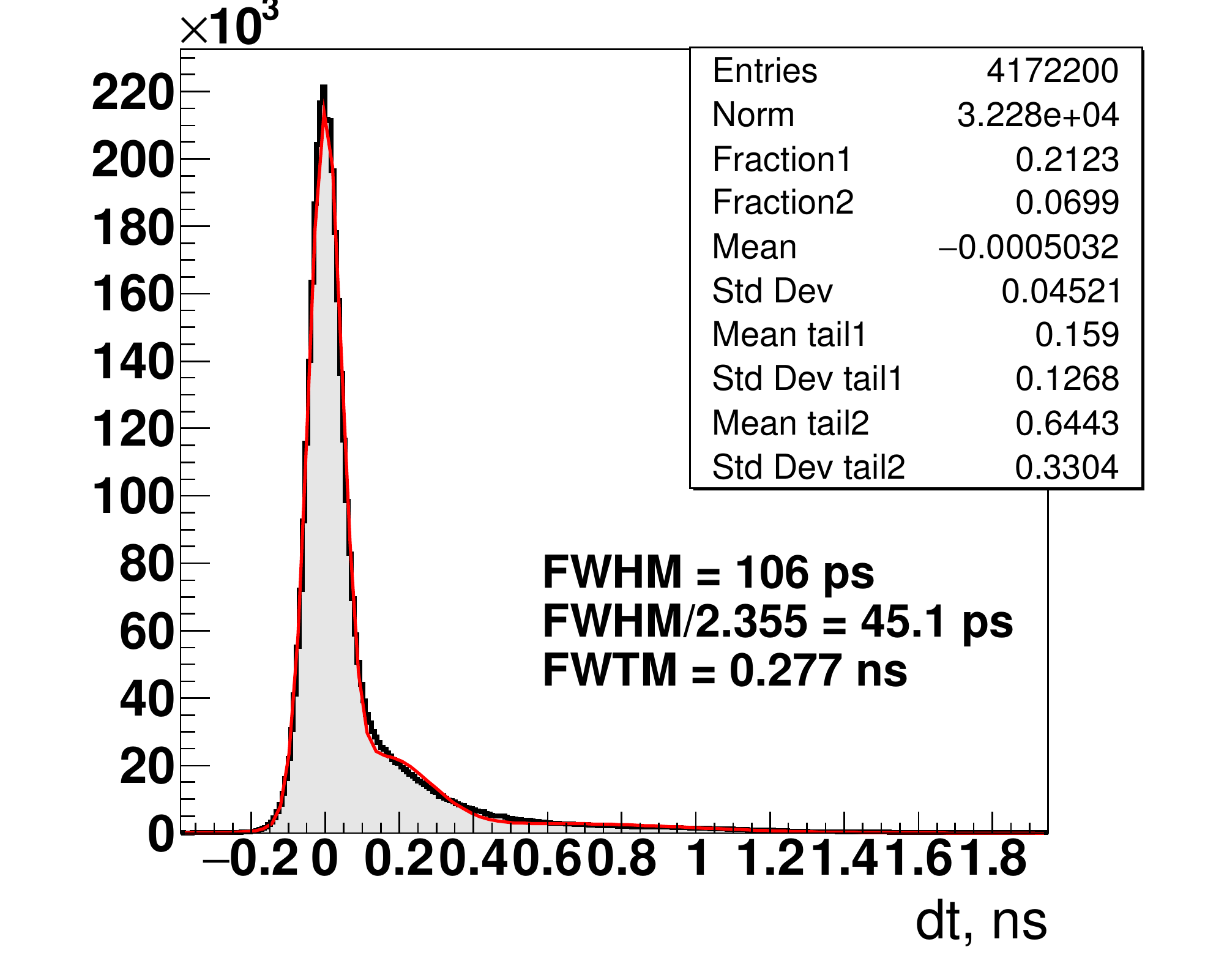}
}
\caption{
Per-channel distribution for time difference between PMT signal and laser trigger 
for two selected channels.
\label{fig:channel_dt}
}
\end{figure}

A 2D distribution of the $t_1$ parameter is shown in Fig.~\ref{fig:mean}. 
We observe an important dispersion of the order of 50--80~ps  inside individual readout channel. 
Over a single anode pad (we connect 4 anodes in one readout channel), this dispersion is smaller, but still present.
In order to obtain the optimal performance from the detector, the signal propagation delays should be calibrated and taken into account.
Unfortunately, the current readout scheme has no means to make the exact correspondence  between an 
event and (x,y) coordinates. 
The dispersion of the delays inside each channel is an important and irreducible limitation for obtaining the optimal performance from 
the device in the current readout design. Fig.~\ref{fig:channel_dt} shows the distribution of time difference per channel. 
The width of the distribution is significantly larger, compared to the per-position distribution (Fig.~\ref{fig:pixel_dt}), 
typically 105 -- 145~ps, see Tab.~\ref{tab:channel_resolution}.
Central channels have better performance due to the smaller dispersion in the $t_1$ parameter. 
Due to the charge sharing effects, we observe the correlation between the mean and the signal amplitude, so for further test 
we apply a calibration of delays ($t_1$ parameter) versus signal amplitude, individually for each channel.

The common cathode signal could be used as a sum of all anode signals, 
for example, for triggering. Due to the large capacity of the cathode,
the signal is slower and smaller than a typical anode signal.
Nevertheless it allows to reach a reasonable resolution  in time with a typical FWHM of 120 -- 180~ps 
for a fixed position. 
The Fig.~\ref{fig:mean_cathode} shows the peak timing (parameter $t_1$ in~the~Eq.~(\ref{eq:reso_func})) for the 
cathode signal as a function of (x,y) coordinates. As one can see, the signal needs around 250~ps to arrive from the 
far end of the electrode to the point of the signal readout.

\begin{table}
\begin{center}
\begin{tabular}{r|c|c|c|c|}
\diagbox{{Channel y}}{Channel x}  & 7 -- 19 mm & 20 -- 32 mm & 33 -- 45 mm & 46 -- 58 mm \\ 
\hline 

48 -- 60 mm & 148  & 137  & 146  & 135  \\ 
35 -- 47 mm & 119  & 108  & 106  & 134  \\ 
22 -- 34 mm & 110  & 115  & 120  & 118  \\ 
9  -- 21 mm & 120  & 132  & 121  & 133  \\ 

\end{tabular}  
\caption{FWHM per channel (ps) for the time difference distribution (see text).
\label{tab:channel_resolution}}
\end{center}
\end{table}

\section{Coincidence Resolving Time}
\label{sec:CRT}
According to Eq.~(\ref{eq:timereso}) we expect to reach the time resolution per detector of about 
50~ps (SD) in the best case, corresponding to the CRT values of 170~ps.
As will be demonstrated in the following, this expectation is optimistic, in particular due to the presence of the non-gaussian tail 
in distributions of the photon trajectories dispersion and the PMT timing resolution.

\subsection{Measurement}
\begin{figure}
\includegraphics[width=.48\textwidth]{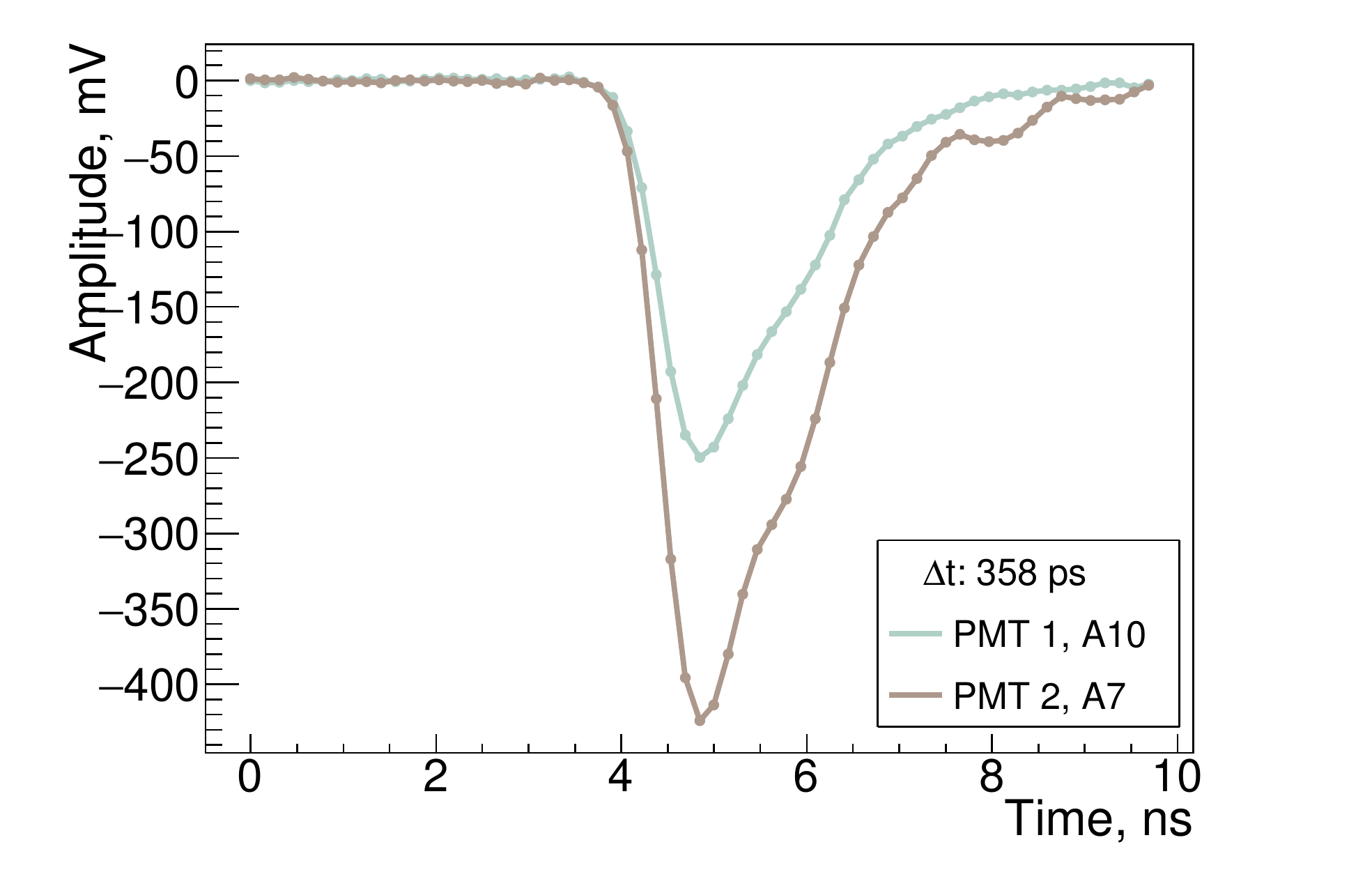}
\hfill
\includegraphics[width=.48\textwidth]{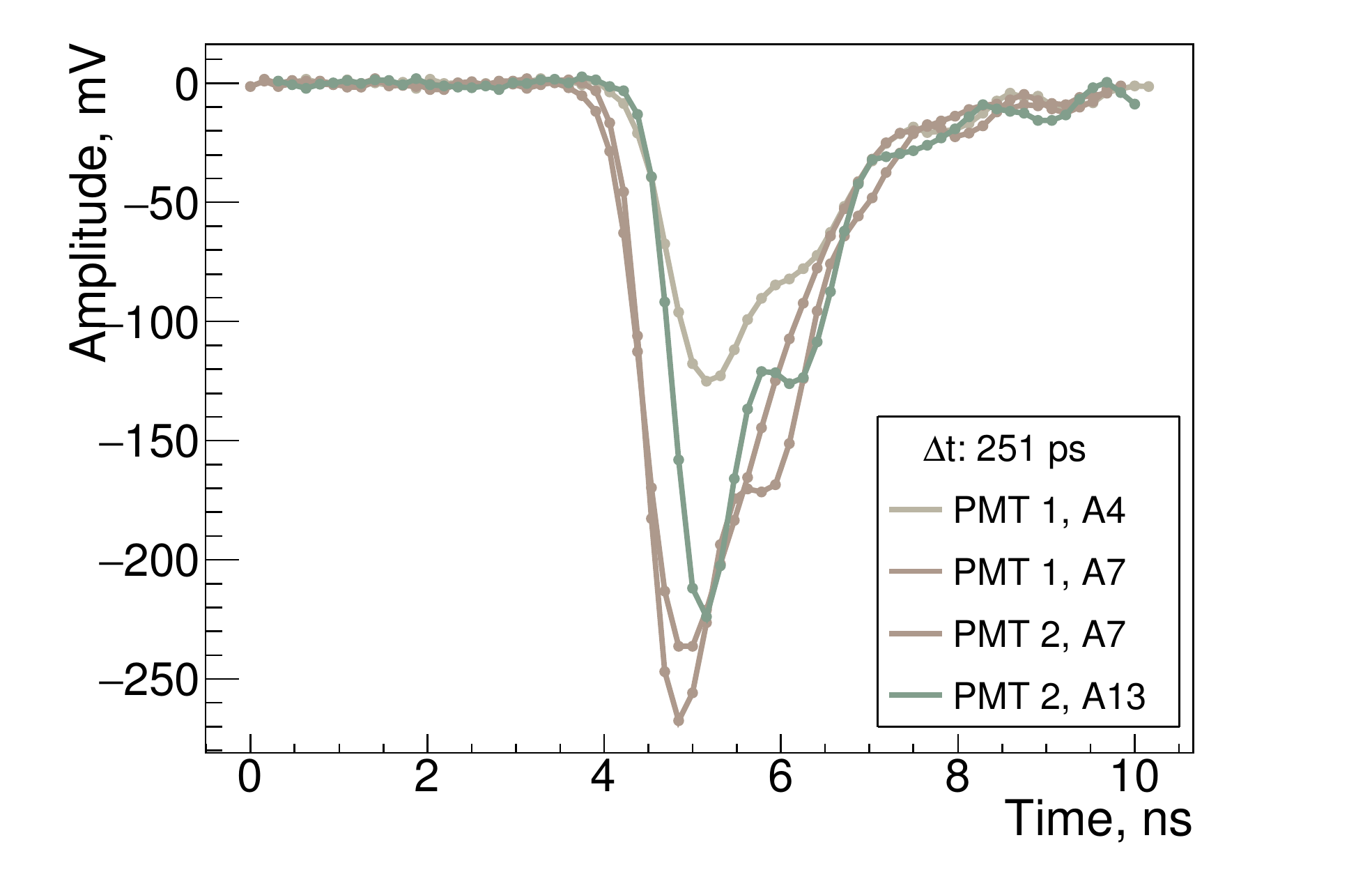}
\caption{Example of signals registered in coincidence by two detection modules.
\label{fig:signals}
}
\end{figure}

To measure the time resolution  we use two detector modules described in the section~\ref{sec:det}. 
Two detectors are installed front-to-front on the optical bench at a distance of 76~cm between them.
In the center, we place a \Na\ radioactive source with a thin, disk-like active area of 10~mm diameter,
encapsulated in 10~mm thick plexiglass disk. With such configuration, the probability of simultaneous detection of
1.27~MeV and 511 keV~photons is less than 0.3\% and the obtained results could be interpreted as the CRT for detecting two back-to-back 
511~keV photons. 
We readout both detectors using a 32-channel SAMPIC module and the coincidence trigger is realized by the module itself.
In particular, we register all events where any anode signal from one PMT is in coincidence with any anode signal from 
another PMT within the time window of 20~ns. 
Typical recorded signals are shown in~Fig.~\ref{fig:signals}.
The data acquisition rate is about 400 coincidences/s. 
The measured random coincidence rate due to the dark count rate of PMTs is 4.7 coincidences/s, but these events are uniformly distributed in the range [-10~ns, 10~ns]
and, hence, represent only a negligible fraction of events, 0.1\%, in the signal region [-1~ns, 1~ns].

\begin{figure}
\includegraphics[width=.48\textwidth]{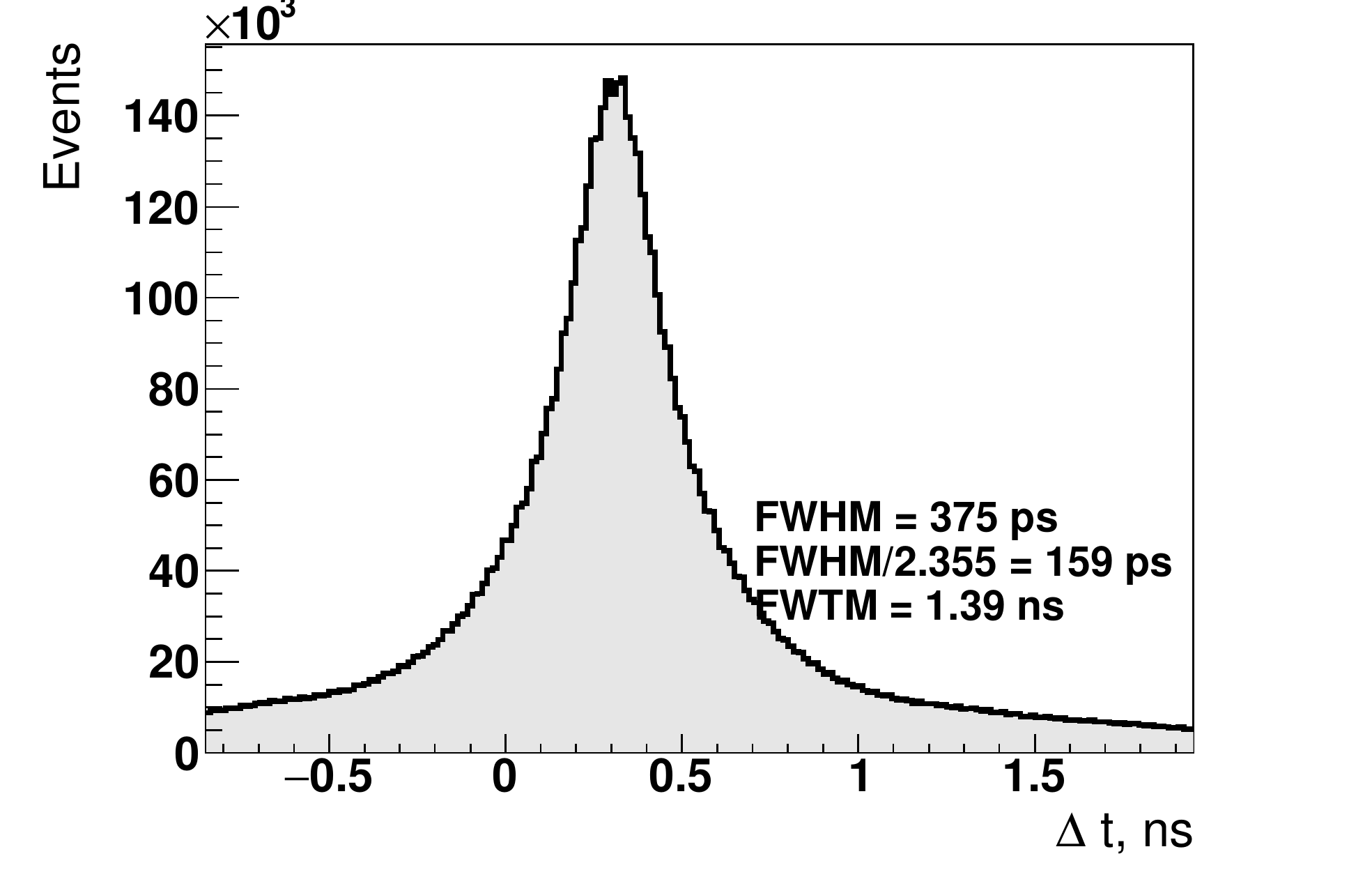}
\hfill
\includegraphics[width=.48\textwidth]{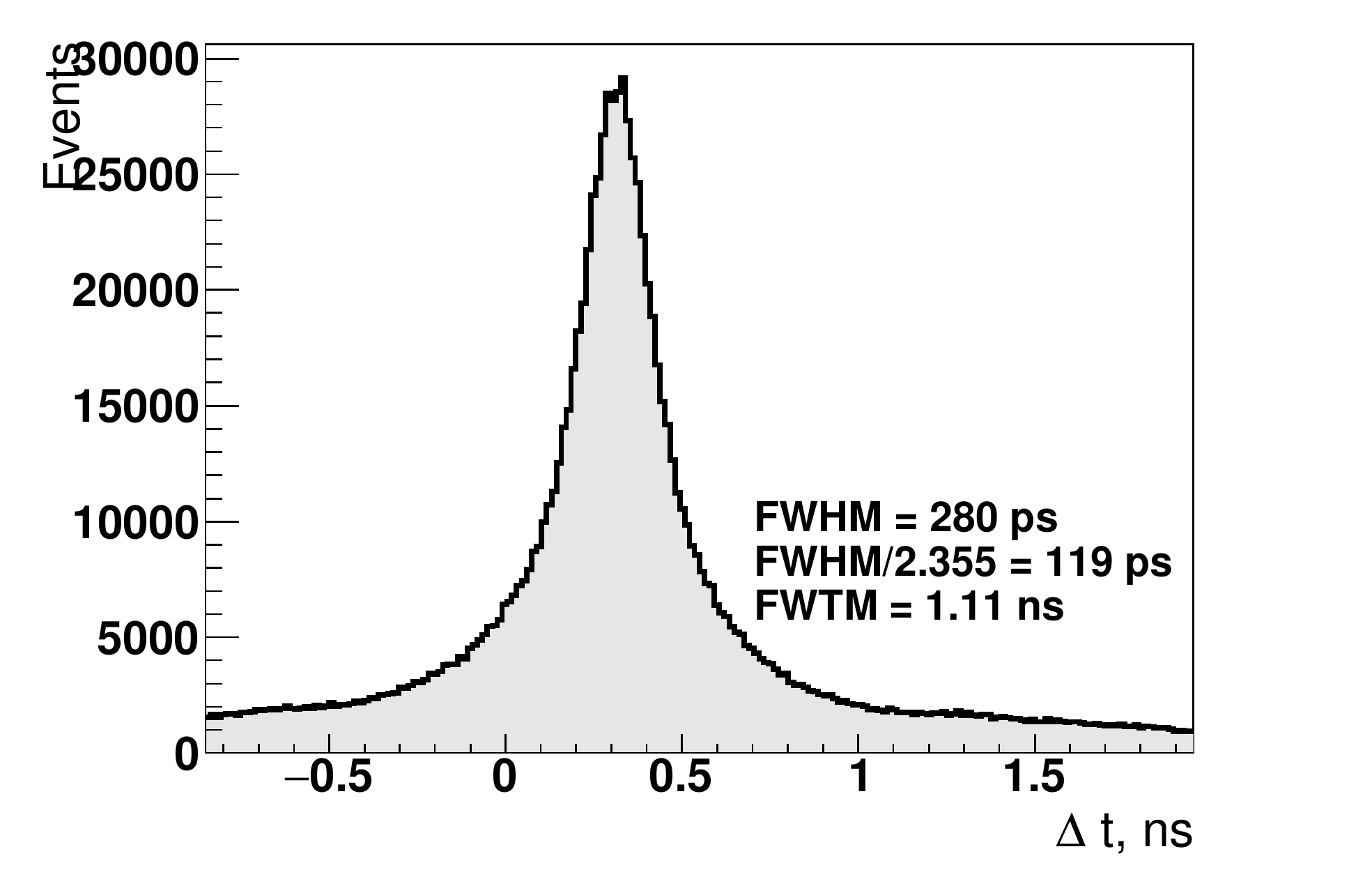}
\caption{Measured difference in time between two detection modules for the full surface (left) or 4 central channels only (right).
\label{fig:time_diff}
}
\end{figure}

\begin{figure}
\includegraphics[width=.48\textwidth]{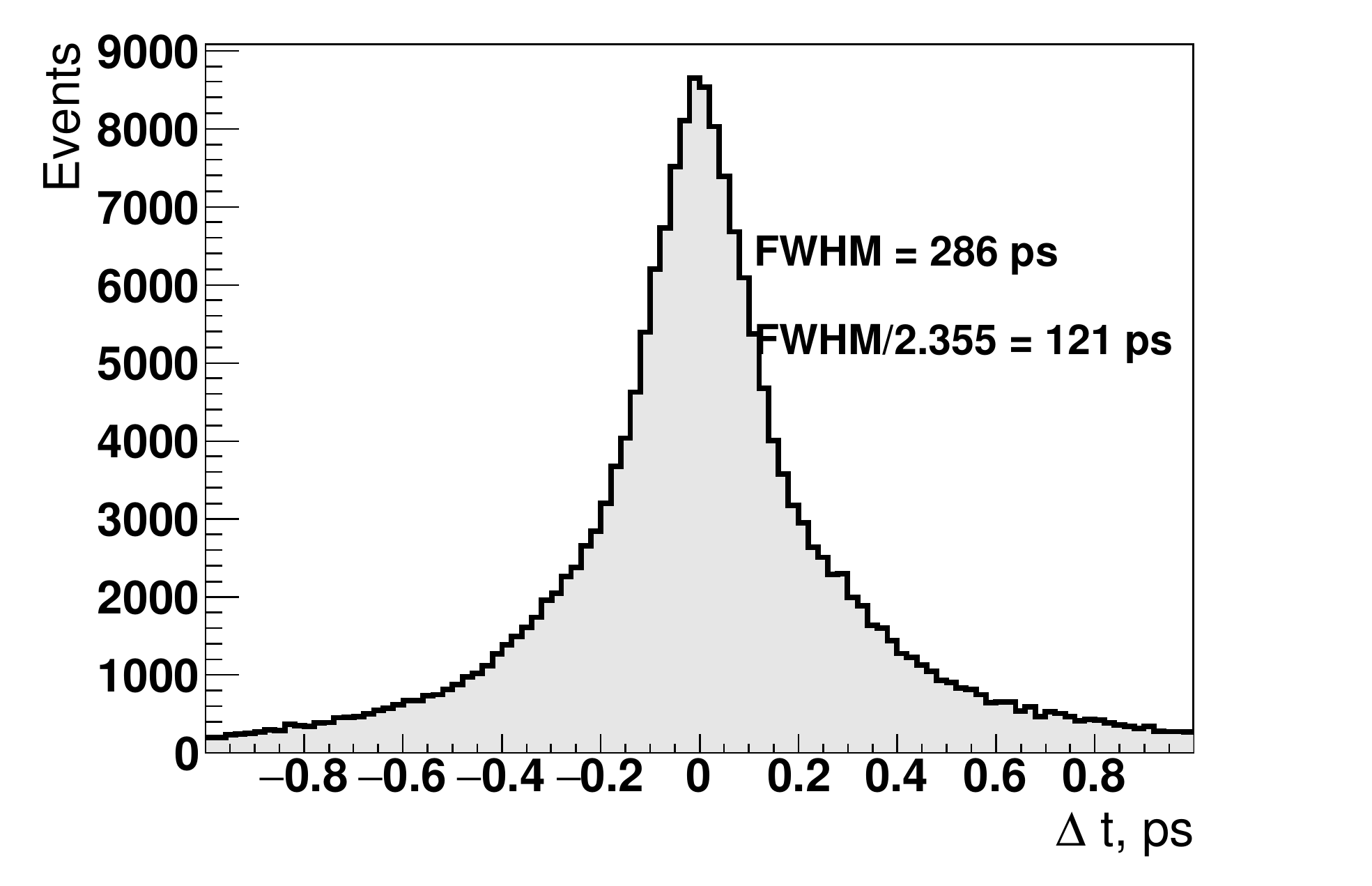}
\hfill
\includegraphics[width=.48\textwidth]{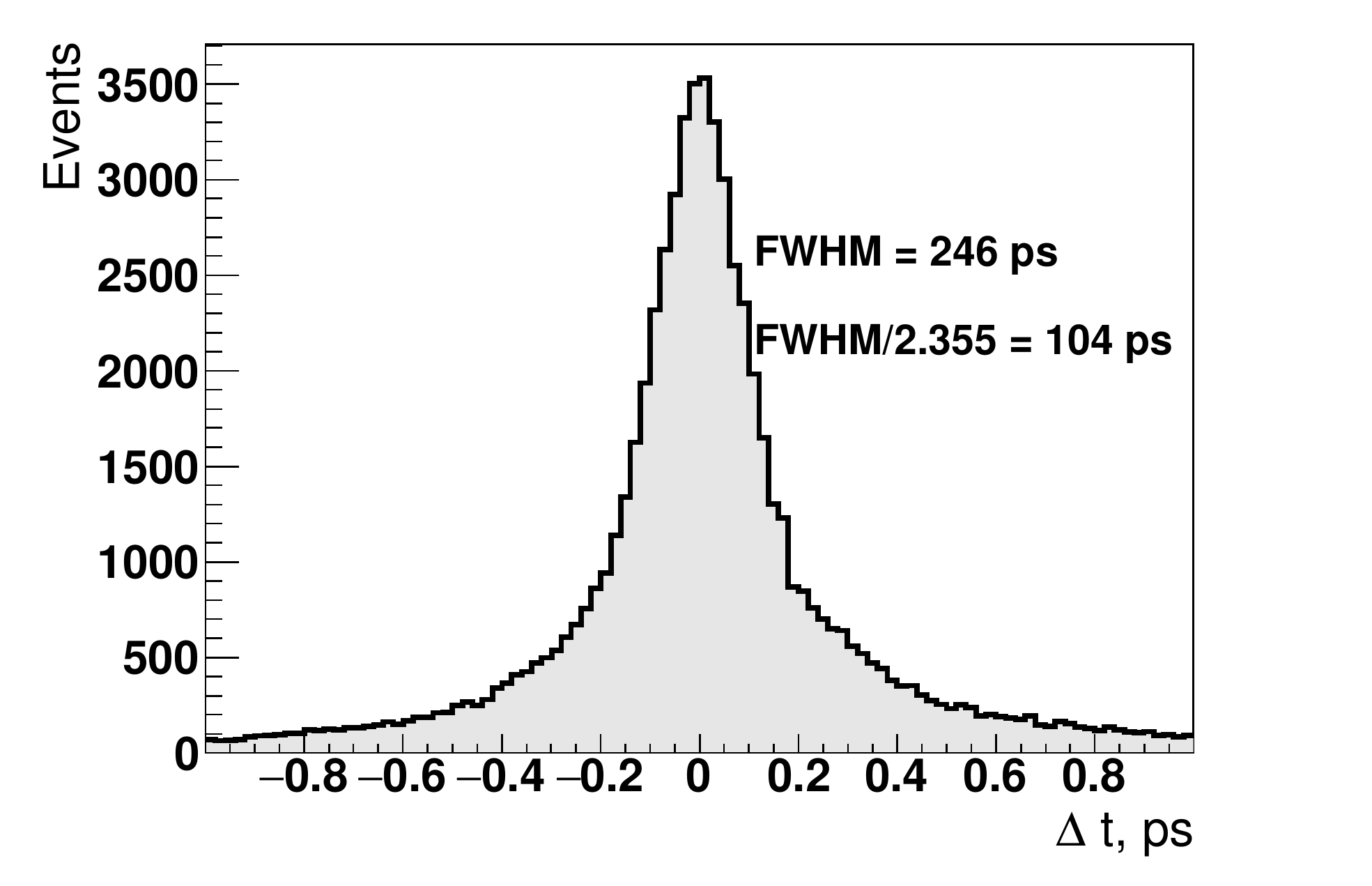}
\caption{Simulated difference in time between two detection modules for the full surface (left) or 4 central channels only (right).
\label{fig:time_diff_sim}
}
\end{figure}

The distribution for the measured difference in time is shown in Fig.~\ref{fig:time_diff}, where we chose 30~mV threshold
for the PMT with the amplification \unit{24}{\db} and 100~mV for the PMT with the amplification \unit{40}{\db}.
As seen on these pictures, the measured CRT is worse for the full surface of the PMT, than for the central channels only.
The difference is caused mainly by the finite size of the radioactive source active area (diameter 10 mm) leading to a
higher efficiency to detect coincidence by the central channels compare to the peripheral ones. 
In consequence, the peripheral channels have large fraction of events triggered
by the photons reflected several times in the crystal and, hence, worse time resolution.

Simulated distributions show 14$\%$ better resolution for the central channels and 
30$\%$ better for the full surface, see Fig.~\ref{fig:time_diff_sim}.
It could be explained by the fact, that quality of the surface has an important role 
for the simulation of the photon reflection inside the crystal.
We used crystals with the surfaces polished to optical quality,
but the actual quality is not measured. 
The results presented in Fig.~\ref{fig:time_diff_sim} are simulated using 
Geant4 UNIFIED model~\cite{Levin1996} assuming the Gaussian distribution of the photon direction
due to reflection from the micro-facets  with a standard deviation of \unit{0.1}{\degree}.
We set the specular lobe probability to one and all others (specular spike,  
backscatter ans Lambertian) to zero. Any difference in surface simulation with the real surfaces quality
will affect mainly the peripheral channels.

\subsection{Discussion}

In this study we demonstrated the possibility to build a Cherenkov based crystalline detector for 511~keV photons. 
Due to high density and high atomic number of the \PbF\ crystal, as well as the large detection surface, 
such a detector provides an efficiency of 24\% suitable for building a TOF-PET scanner. 
For example, using the crystal matrix made with 6x6x10~mm${^3}$ \PbF\ crystals attached to the MCP-PMT, one can design the whole body 
Cherenkov PET scanner, with  performance comparable to conventional scanners, as estimated by the simulation 
with somewhat optimistic hypothesis \cite{Alokhina2018, Alokhina2018a}.

\begin{figure}
\subfloat[Without tail in the PMT resolution function.
\label{fig:dt_mc_notail}] {
\includegraphics[width=.48\textwidth]{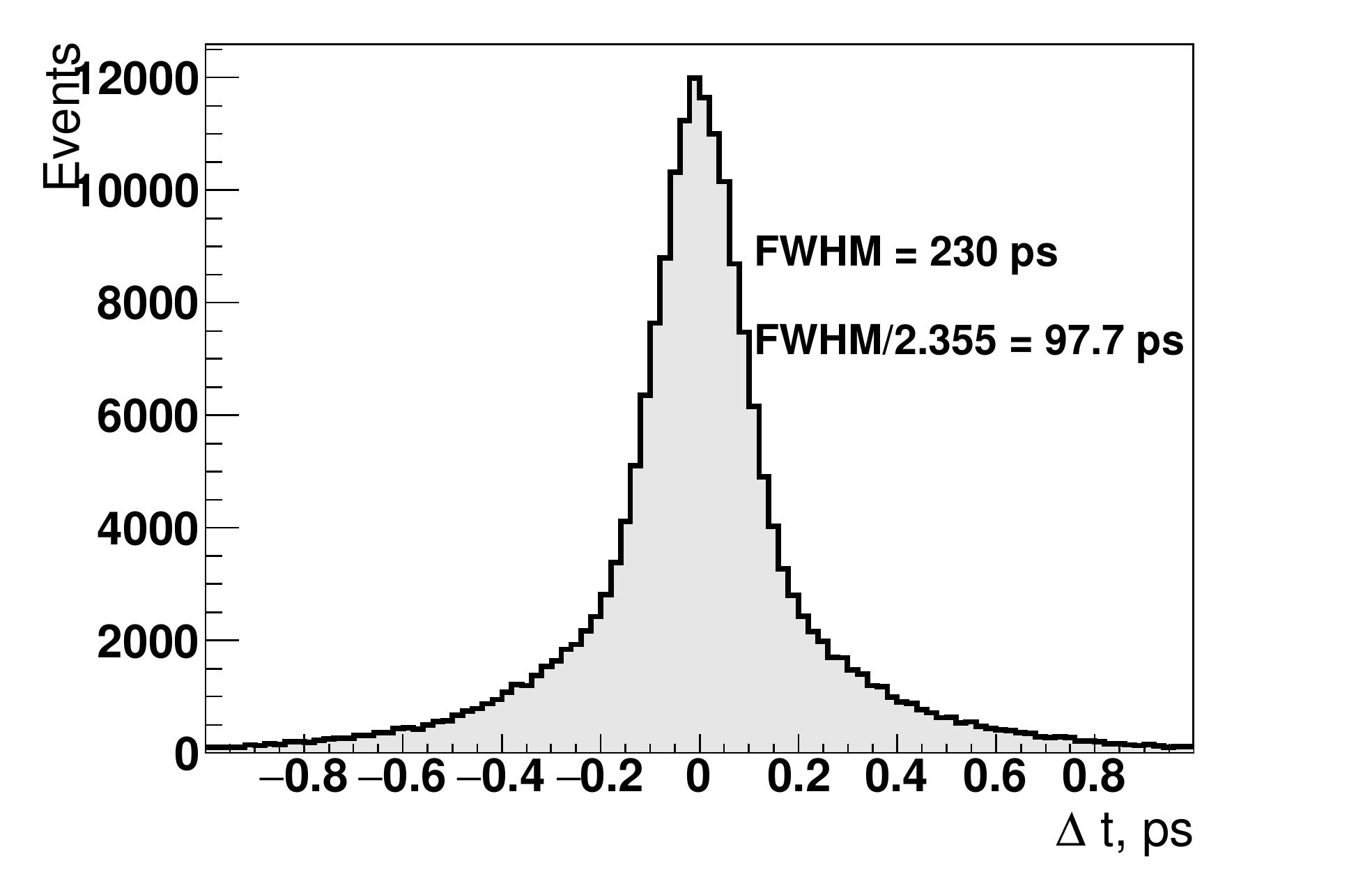}
}
\hfill
\subfloat[No degradation from the dispersion of signal delays inside channel.
  \label{fig:dt_mc_pixel}]{
  \includegraphics[width=.48\textwidth]{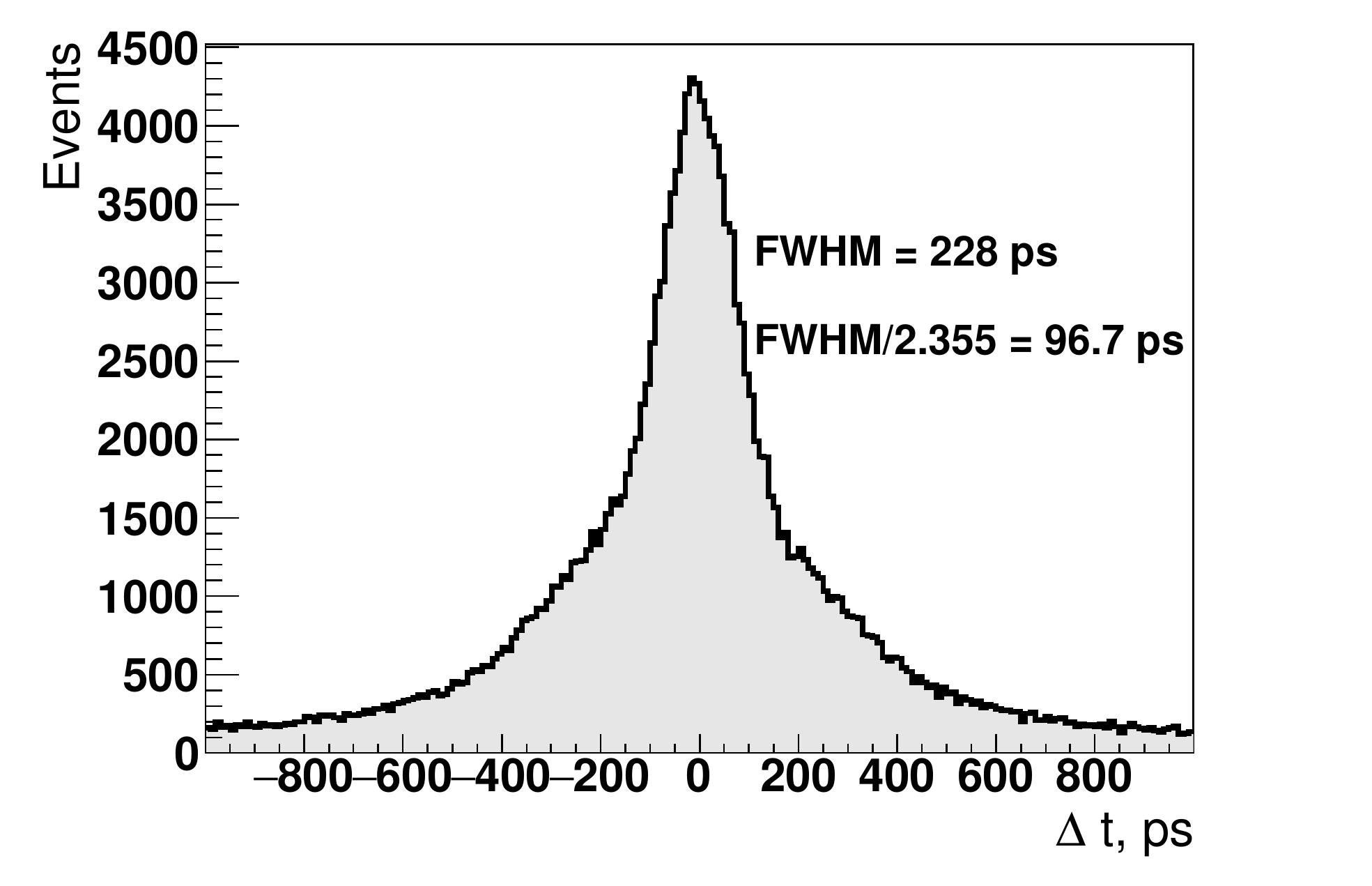}
}
\caption{Simulated time difference between two detection modules with modified PMT resolution function.}
\end{figure}

Nevertheless, the performance of such a detection module stays modest due to the several limitations.
First of all, the obtained CRT exceeds a lot the gaussian expectation $\sqrt{2}\sigma_{\text{detector}}$ due to the presence
of non-gaussian tails in the distributions. These tails are due to the dispersion of the photon trajectories and especially due to the MCP backscattered electrons, which 
generate delayed signals for at least 25\%  of events in the range of 100 -- 2000~ps (for Planacon XP85012).
Any reduction in the fraction of such events will improve significantly the CRT.
For example, in Fig.~\ref{fig:dt_mc_notail} 
we shows the simulated distribution of the time difference between two detection module, where we remove the
tail in PMT resolution function by assigning parameters $f_1$ and $f_2$ from eq.~\ref{eq:reso_func} to zero. 
This distribution has smaller width by about 20\% 
compare to the  Fig.~\ref{fig:time_diff_sim}. 

As mentioned in the section~\ref{sec:pmt_reso}, the dispersion of the signal delays inside one channel requires the use of small pads individually readout,
but small pads increase the fraction of events affected by the charge sharing effects. Ideally, a continuous  readout with 
the possibility to reconstruct $x$ and $y$ position of each photon allows to calibrate and correct for delays, 
so improving the time resolution. 
Fig.~\ref{fig:dt_mc_pixel} shows the distribution of the time difference between two detection module, 
assuming no degradation from the signal delays inside channels. 
As expected, the CRT is improved by 20\% compared to Fig.~\ref{fig:time_diff_sim}.

The main disadvantage in using the  Cherenkov radiation compare to the scintillation 
is the small number of generated photons and detected photoelectrons.
An increase in the number of  photoelectrons generated at the photocathode will improve very significantly the time resolution.
For example, Planacon XP85012 PMT features a bi-alkali photocathode with a maximum efficiency of 22\%~\cite{XP85012_DataSheet}, 
but better photocathodes with efficiency up to 30\% are available now~\cite{Orlov2016Apr}. 
An important loss in efficiency is also caused by the non-ideal optical interface as described in the section~\ref{sec:eff} and 
reflection of Cherenkov photons with a large incident angle from the crystal border.
Increasing the critical angle would increase the number of the detected optical photons.
In addition, it reduces the number of photon reflections in the crystal, and thus will improve the time resolution.
To increase the critical angle, one needs to increase the refraction index of the external media. The conventional way to do so is to 
apply the optical media between PMT and crystal. Unfortunately, it is not possible to find an optical media with high refraction index,
and, simultaneously, transparent in the deep UV region, required for the detection of Cherenkov photons.
An alternative way to improve the optical interface, proposed by us previously~\cite{bonding_patent},
is to use a molecular bonding between crystal and PMT window. 
This procedure glues together PMT window and crystal without using any intermediate media. If the PMT window 
has a high refractive index (e.g. sapphire), it will significantly improve the quality of the optical interface.   
Such an operation requires polishing of both surfaces to the roughness less than 1~nm and
planarity below 1~\um. In addition, both surfaces should be free of dust particles or contamination, especially hydrocarbons, 
see e.g.~\cite{Moriceau2012331}. This technique was judged to be too challenging, 
especially taking into account the necessity prepare a photocathode on the bounding object under the ultra-high vacuum and  
high temperature.

Finally, we decided to improve the optical interface by removing completely  the border between PMT window and 
a Cherenkov radiator and depose photocathode directly on the crystal. 
We choose to implement this technique using the PbWO$_4$ crystal, 
which is almost as good Cherenkov  radiator as \PbF\ and, in addition, produces a small number of fast scintillating photons. 
This idea is the main element of our future project named ClearMind~\cite{patent_ClearMind}.

\section{Conclusion}
In this paper we studied the possibility to construct a Cherenkov PET detection module with high efficiency 
and good timing performance using \PbF\ crystal and commercial MCP-PMT. 
We measured an efficiency of 24\% to detect a 511~keV photon in a 10~mm thick crystal. 
This value is sufficiently high to be used in PET if high TOF resolution  is reached.  
The use of SAMPIC fast digitization module allows to minimize the electronics contribution to the time resolution
to the level below 20~ps (FWHM) and provides the high rate readout capability of up to $10^5$ events/s.
We developed the fast scanning system to calibrate the time response of the PMT and used it for precise calibration of 
Planacon XP85012. We observe a good time response for the entire PMT surface with the resolution of about 90~ps (FWHM) 
 and the presence of delayed events in the range of 100 -- 2000~ps  at the level of 25\%.

Finally, we measured the CRT between two identical modules of about 280~ps,
limited by the low number of detected optical photons, the PMTs performances and the implemented readout scheme.
The time resolution, reachable with the proposed approach, limits the potential of such technique for full-size scanner.

We are working on an improvement of the detection module performance by improving the optical interface between PMT window and 
crystal and by improving the PMT readout scheme.

\acknowledgments
We thank the Photonis representative Serge Duarte Pinto for constructive discussions on MCP-PMT technology. 

We acknowledge the financial support by the LabEx P2IO R\&D project of the region Ile-de-France, 
the "IDI 2015" project funded by the IDEX Paris-Saclay, ANR-11-IDEX-0003-02, the PhD financial support by the French
embassy in Ukraine, 
DRI University Paris-Saclay and the Programme Transversal Technologies pour la Sant\'e, of CEA. 
This work is conducted in the scope of the IDEATE International Associated Laboratory.

\bibliography{References}

\end{document}